\documentclass[12pt ]{article}

\usepackage{lineno}
\usepackage[hidelinks]{hyperref}
\usepackage{amsmath,amssymb,amsfonts,bm}
\usepackage{algorithmic}
\usepackage{graphicx}
\usepackage{textcomp}
\usepackage{xcolor}
\usepackage{subcaption}
\usepackage{mathtools}

\usepackage{setspace}
\usepackage{multicol}
\usepackage{algorithmic}
\usepackage{graphicx}
\usepackage{textcomp}
\usepackage{mathtools}
\usepackage{xcolor}
\usepackage{subcaption}
\usepackage{mathrsfs}
\usepackage{caption}
\usepackage{float}
\usepackage{indentfirst}
\usepackage{blindtext}
\usepackage{dblfloatfix}
\usepackage{balance}
\usepackage[a4paper, total={6.5in,9in}]{geometry}
\usepackage{flushend}
\usepackage{commath}
\usepackage{bbold}
\usepackage{hyperref}
\usepackage{mathbbol}
\usepackage{comment}
\usepackage{blindtext}
\usepackage[font=small,labelfont=bf]{caption}
\usepackage{subcaption}
\usepackage{graphicx}
\begin{document}
\begin{center}
\Large{{\noindent Accurate PET Reconstruction from Reduced Set of Measurements based on GMM}}
\end{center}
\begin{center}
    Tomislav Matuli\'{c}*, Damir Ser\v{s}i\'{c}
\end{center}
\begin{center}
    \small{{\noindent * Corresponding author}}
\end{center}
\begin{center}
Department of Electronic Systems and Information Processing, University of Zagreb Faculty of Electrical Engineering and Computing, Unska 3, 10000 Zagreb, Croatia
\end{center}
\noindent\rule{\textwidth}{1pt}
\begin{abstract}
In this paper, we provide a novel method for the estimation of unknown parameters of the Gaussian Mixture Model (GMM) in Positron Emission Tomography (PET). A vast majority of PET imaging methods are based on reconstruction model that is defined by values on some pixel/voxel grid. Instead, we propose a continuous parametric GMM model. Usually, Expectation-Maximization (EM) iterations are used to obtain the GMM model parameters from some set of point-wise measurements. The challenge of PET reconstruction is that the measurement is represented by the so called lines of response (LoR), instead of points. The goal is to estimate the unknown parameters of the Gaussian mixture directly from a relatively small set of LoR-s. Estimation of unknown parameters relies on two facts: the marginal distribution theorem of the multivariate normal distribution; and the properties of the marginal distribution of LoR-s. We propose an iterative algorithm that resembles the maximum-likelihood method to determine the unknown parameters. Results show that the estimated parameters follow the correct ones with a great accuracy. The result is promising, since the high-quality parametric reconstruction model can be obtained from lower dose measurements, and is directly suitable for further processing.
\end{abstract}
\noindent\rule{\textwidth}{1pt}

\section{Introduction}
Positron emission tomography is a medical imaging modality that measures metabolic activity of the observed tissue. It is based on electron-positron annihilation. The annihilation happens due to the radioactive $\beta^+$ decay of the radioactive tracer that is injected into the measured tissue. The result of the annihilation is two high-energy photons (511\,keV each) that travel along the same line but in opposite directions. If two photons hit detectors in a short-coincidence time-window, then the data is being stored and it is considered as a valid measurement event. Virtual path between the two detectors is called line of response (LoR).  The main challenge in medical imaging is to reconstruct image from such measurement data. In PET imaging systems, a list of all coincidence events corresponds to a measurement data set.\cite{Vaquero2015}\cite{Cherry}

Analytical reconstruction algorithms \cite{Panetta}\cite{Defrise1998}\cite{Kinahan1989}\cite{Matulic2021} (filtered back-projection, back-projection filtering) are rarely used nowadays. They are fast and computationally inexpensive, but often do not include information about the real PET imaging systems and, therefore, produce images of lesser quality. Iterative methods are the golden standard in PET image reconstruction.\cite{Wang2019} \cite{Zhang2001a} The core of iterative algorithms is the system matrix, which represents the PET imaging system.\cite{Khateri2019}\cite{Leahy2000} Maximum-likelihood expectation-maximization (MLEM) algorithm \cite{Shepp1982} is the most widely used iterative method. A major drawback of the MLEM algorithm is its slow convergence. To accelerate the convergence, the ordered subset expectation-maximization (OSEM) was introduced in \cite{Hudson1994}. It significantly reduces the time needed for image reconstruction (more subsets = faster convergence), but it does not guarantee the maximum likelihood solution. More recently, deep neural networks are exploited in modeling of the PET imaging systems \cite{Han2017}\cite{Berker2018}\cite{Wang2018} and for image reconstruction \cite{Gong2019}\cite{Xie2020}\cite{Hggstrm2019}. Pay attention that all mentioned algorithms result in a spatially-discrete reconstruction model, namely, a grid of pixels or voxels.

The Gaussian mixture model \cite{Reynolds2015} (GMM) is a well-investigated approach in a variety of classification and segmentation problems.\cite{Friedman1997a} \cite{Ralasic2018abc} The application of the GMM can be found in many problems in biometrics, signal processing, and speech modeling. \cite{Reynolds2015} \cite{Reynolds2000} \cite{Majumder2022} \cite{Yu2011} \cite{Leger2011} \cite{Li2018} \cite{Tafro2022} \cite{Raitoharju2020} Two different ways are often used for the estimation of parameters of GMM: the expectation-maximization method \cite{Hastie2009}\cite{Murphy2012abc} and the method of moments \cite{Kane21ab}\cite{Khouja2022}. Method of moments relies on tensor moments of higher order to estimate the unknown parameters of the GMM. The theory of such a method is well known and thoroughly investigated. The major drawbacks are:
\begin{itemize}
\item high-order moments are computationally inefficient \cite{Rahmani2020}, since the $k$-th moment of an $n$-dimensional random variable is a tensor of size $n^k$;
\item the method of moments leads to a multivariate polynomial system. Statistically meaningful solutions of a such system often do not exist or are not unique.\cite{Wu2020}
\end{itemize}
A minimum order of the moment needed for the estimation of unknown parameters increases with the number of components in the Gaussian mixture.
Hence, Expectation-Maximization (EM) algorithm is studied more intensely than the method of moments. Although the EM algorithm converges slowly \cite{pmlr-v23-anandkumar12}, its drawbacks are far less concerning. Notice that the obtained GMM model is spatially-continuous, and virtually of infinite resolution. It motivates our research.

In this paper, we present a new method for estimation of the GMM parameters in 2D PET imaging. We can state this problem in another way - how to estimate the unknown GMM parameters from the lines of response that originate from some point that follows Gaussian distribution, but the line is fired under an arbitrary angle? All the methods mentioned in the previous paragraph deal with point-wise samples from the GMM distribution itself, while we deal with the lines that originate from those point-wise sources. To the best of our knowledge, the mentioned issue is unexplored or under-explored \cite{Tafro2022}, \cite{Koscevic2021}, and this paper gives a novel insight into it. 

In this work, the problem and its solution is set in the projection domain. Each point in the image gives a sine function in the projection domain. Thus, the projection domain is often called sinogram: a collection of sines. Vice-versa, a single point in the projection domain corresponds to a line of response in the image. Notice that centers of each Gaussian component are points in the image domain, too. 

In our work, we exploit the following facts:
\begin{enumerate}
\item the mean vector (i. e. the center $\bm\mu$) corresponds to the sine function in the projection domain;
\item projection under an arbitrary angle of a bivariate normal distribution is a univariate normal distribution. This is a special case of the marginal distribution theorem of the multivariate normal distribution;
\item expressions for calculating higher moments of the Gaussians are well known, and have shown to be useful for estimation of the GMM parameters.
\end{enumerate}
In Section 3 we exploit the mentioned facts, and using some tricks arrange it to get a solvable system of equations. Then, a two-step iterative process that resembles the EM algorithm is exploited for determining all of the unknown parameters of the GMM. The first step updates membership probabilities between the lines of response and the GMM components. The second step reevaluates the parameters of each Gaussian component according to the updated memberships.

Resulting GMM model has several advantages over the usual pixel or voxel based approach. Essentially, it is an infinite resolution continuous model obtained from a reduced set of measurements. Widespread digital processing methods are based on difference equations that approximate the underlying differential equations. In our case, we can apply them directly on the GMM model without approximations.

This paper is divided into six sections. In Section 2 we explain a 2D PET imaging system. We present an estimation method of the mean vector and the covariance matrix for one component of the GMM in Section 3. In Section 4 we show the proposed iterative, EM-like algorithm for the estimation of the unknown parameters of the whole GMM model. The results are given in Section 5. Finally, in Section 6 we conclude the paper.

\section{Two-Dimensional PET Imaging}
\begin{figure}
\centering
\includegraphics[width=0.55\linewidth]{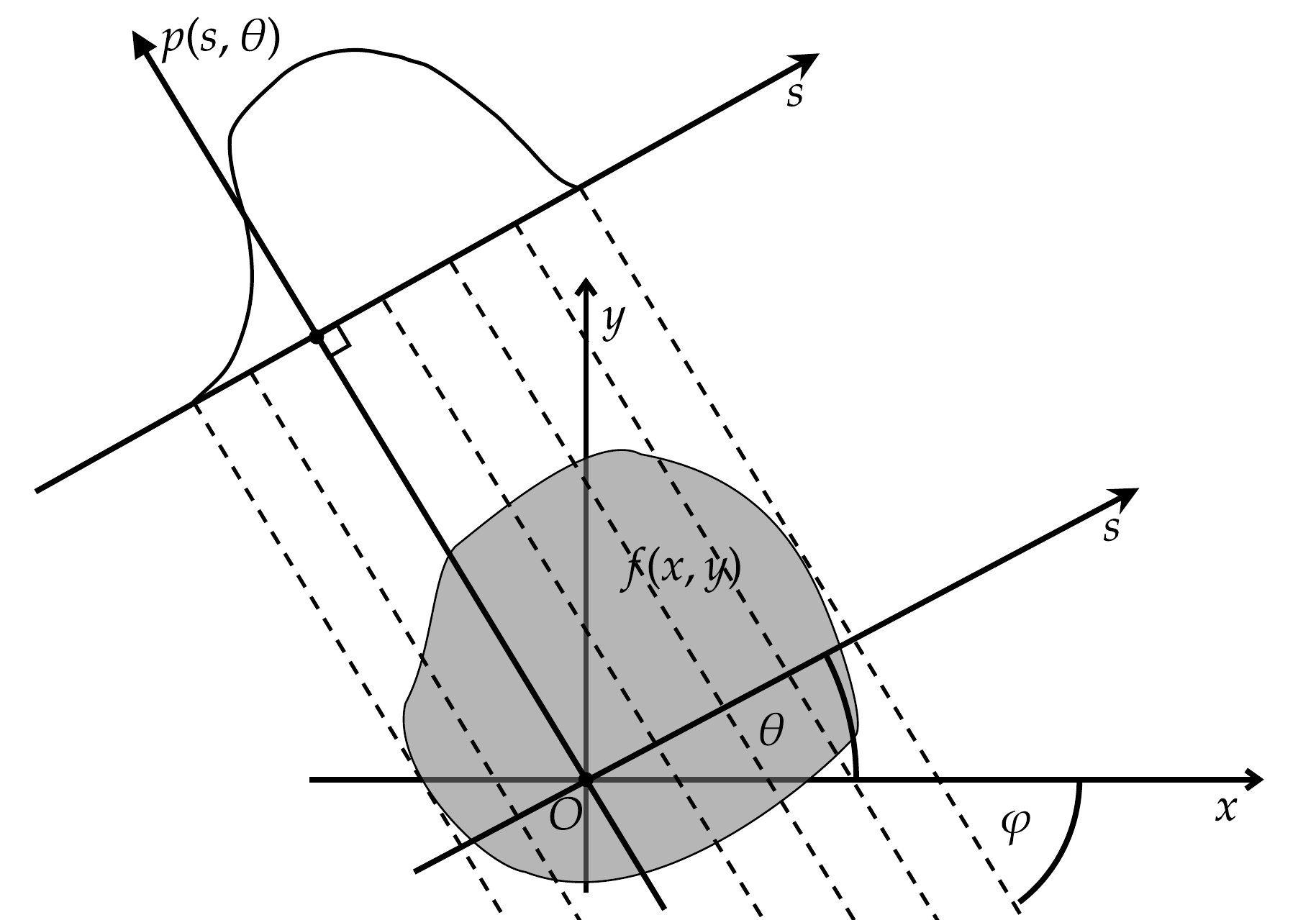}
\caption{Radon transform} 
\label{fig:radon}
\end{figure}

The mathematical background of PET imaging is based on the Radon transform. The expression for the Radon transform is
\begin{equation}
    p(s,\theta)  =
    \displaystyle \int\displaylimits_{-\infty}^{+\infty} \int\displaylimits_{-\infty}^{+\infty} f(x,y)\, \delta(x\cos(\theta) +y\sin(\theta)-s)\, dxdy,
\end{equation}
where function $f(x,y)$ represents metabolic activity of the object that is being scanned by the PET system, and $p(s,\theta)$ models projection at the angle $\theta$ - an angle between the projection line and the x-axis. Value $p(s,\theta)$ corresponds to the line integral that is perpendicular to the projection line, and its distance from the origin is $s$. Described setup is illustrated in Fig. \ref{fig:radon}.

\begin{figure}
\centering
    \begin{subfigure}{0.4\textwidth}
    \includegraphics[width=\linewidth]{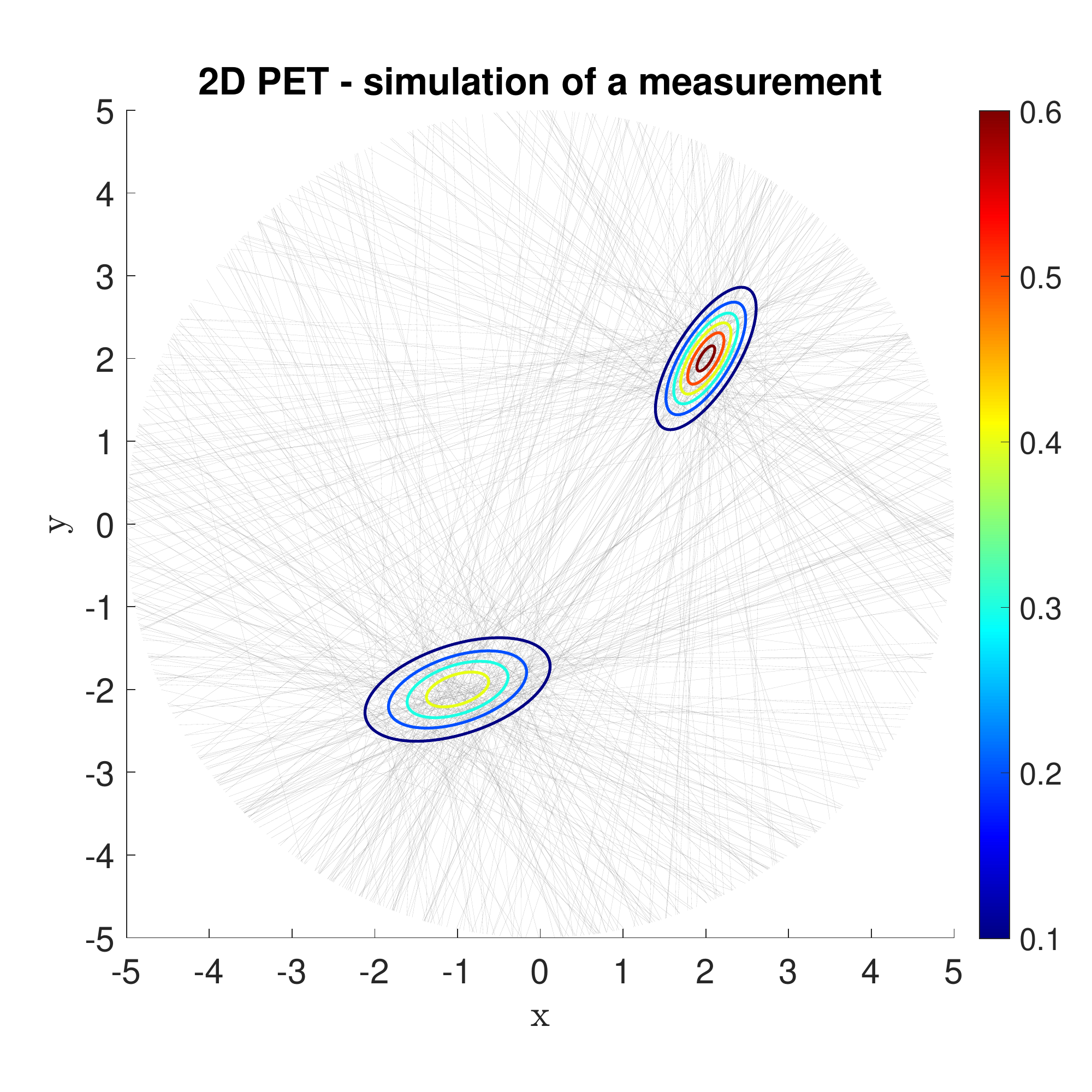}
    \caption{2D PET measurement}
    \label{fig:pet_measure}
    \end{subfigure}
    \begin{subfigure}{0.55\textwidth}
    \includegraphics[width=\linewidth]{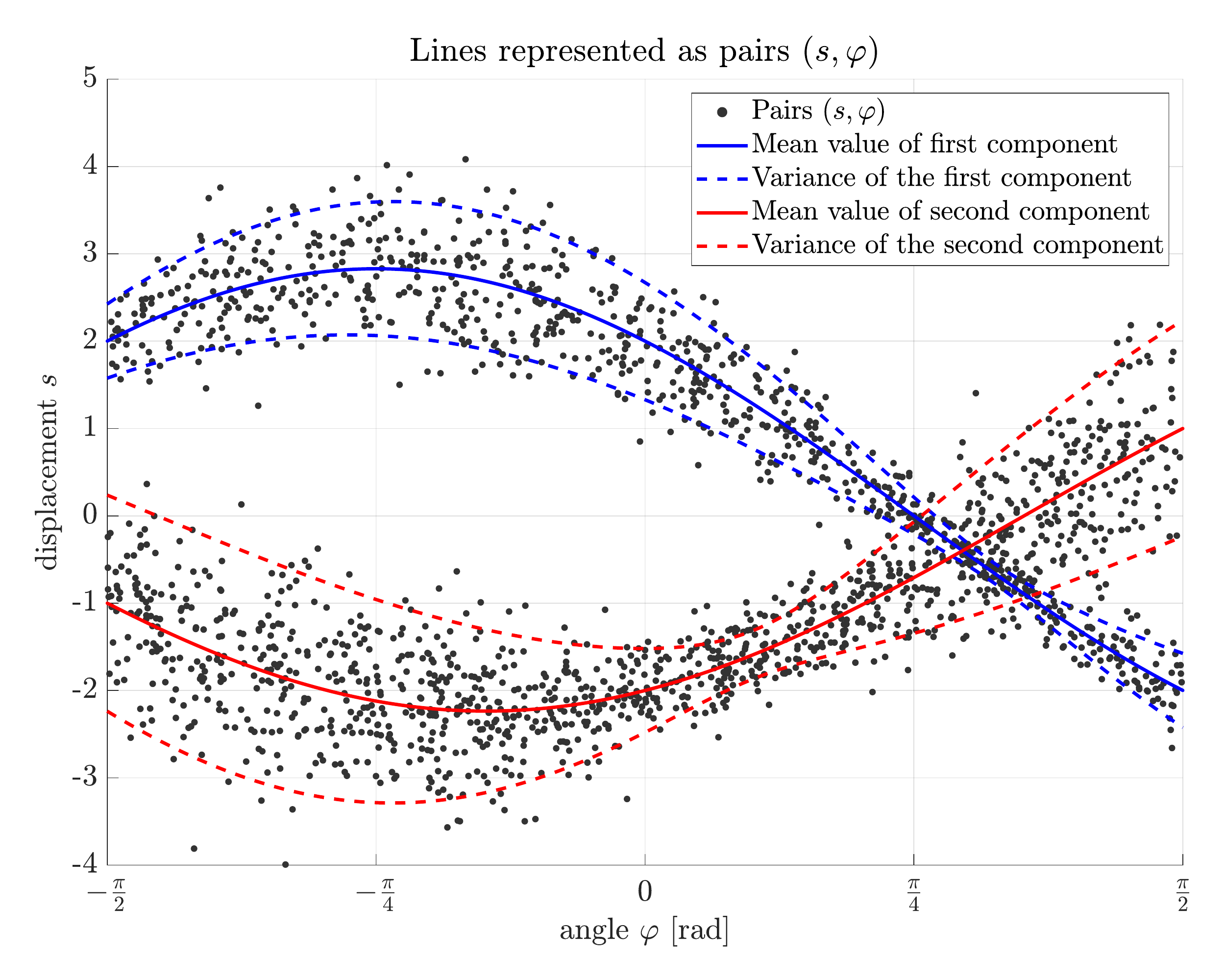}
    \caption{\centering LoRs represented as points $(s,\varphi)$. 
    Solid curves depict means, while dashed curves depict variances, both functions of angle $\varphi$.}
    \label{fig:Sinogram}
    \end{subfigure}
\caption{2D PET measurement} 
\label{fig:PET}
\end{figure}

In our experiment, function $f(x,y)$ is regarded as the probability density function (PDF) of some GMM. Each sample of the GMM is a position of electron-positron annihilation. The samples fire lines of response under an uniformly distributed random angle. Actual positions of annihilation (samples) are latent (hidden). In Fig. \ref{fig:pet_measure}, we can see a simulation of the PET measurement. We generated $N=2000$ lines firing in both directions from some hidden point-wise samples that follow the GMM distribution. Contours are used to indicate the two Gaussian components, and only a subset of LoR-s is displayed for better visibility.

A line of response can be represented as a pair $(s,\varphi)$, where $s\in\mathbb{R}$ is the oriented distance from the origin (see $s$-axis in Fig. \ref{fig:radon} or Fig. \ref{fig:mean_estim}), and $\varphi\in\left[-\frac{\pi}{2},\frac{\pi}{2}\right]$ is the angle between the line and the x-axis. LoR-s correspond to points $(s,\varphi)$ in the projection domain. The integral line and projection line are perpendicular, i.e. the angles $\theta  = \varphi \pm \frac{\pi}{2}$. Measurement is illustrated as a set of points $(s,\varphi)$ in Fig. \ref{fig:Sinogram}, or as a set of lines of the response in Fig \ref{fig:pet_measure}. The mean vector of the Gaussian component traces a sine function in the projection domain, shown as a solid curve. Dashed curves stand for $\pm 3 \sigma$-neighborhood around the mean. 

In the next section, we focus our attention on how to estimate the mean vector $\bm\mu$ and covariance matrix $\bm\Sigma$ of a single component in the mixture. Estimation of the mean vector is done by fitting \emph{the best} sinusoid on pairs $(s,\varphi)$. Covariance matrix $\bm\Sigma$ is estimated from the moments of the probability density function.

\section{Estimation of Parameters of the Gaussian Mixture Model}
The Gaussian mixture model with $K$ components is given by
\begin{equation}
g(\bm x; \tau_k,\bm\mu_k,\bm\Sigma_k)=\displaystyle \sum\displaylimits_{k=1}^K\tau_k f_G(\bm x; \bm \mu_k, \bm \Sigma_k),
\label{eq:GMM}
\end{equation}
where $\bm\mu_k$ is the mean vector, $\bm\Sigma_k$ is the covariance matrix and $\tau_k$ is the weight, all associated to the $k$-th component. In general, function $f_G(\bm x; \bm \mu_k, \bm \Sigma_k)$ is a $d$-variate normal (Gaussian) distribution:
\begin{equation}
f_G(\bm x;\bm\mu_k,\bm\Sigma_k)=\frac{1}{\sqrt{(2\pi)^d|\Sigma_k|}}\exp(-\frac{1}{2}\left(\bm x - \bm\mu_k\right)^\intercal\bm\Sigma_k^{-1}\left(\bm x - \bm\mu_k\right)).
\label{eq:multivar_gauss}
\end{equation}

Weights $\tau_k >0$ should suffice the condition $\sum\displaylimits_{k=1}^K\tau_k = 1$, since $g$ defines the probability density.

Now, we focus only on one component in the mixture. Since we deal with 2D PET imaging, we have bivariate ($d=2$) normal distribution. Also, we assume that a number of components $K$ in the mixture is known in advance. Estimation is possible, but it is beyond the scope of this paper.

\subsection{Estimation of Mean Vector}

First, we take a look of the univariate normal distribution. Let $(x_i)_{i=1}^N$ be the samples of some univariate normal distribution. Then, the least square problem for
\begin{equation}
L_{uni}(c;x_i) = \displaystyle\sum_{n=1}^N (c-x_i)^2
\label{eq:uni_ls}
\end{equation}
has the solution
\begin{equation}
c_{LS} = \frac{1}{N} \displaystyle\sum_{n=1}^N x_i \quad = \hat{\mu}.
\label{eq:uni_ls_sol}
\end{equation}
Obviously, least square solution $c_{LS}$ corresponds to the known mean value estimate of the univariate normal distribution $\hat{\mu}$.
\begin{figure}
\centering
    \begin{subfigure}{0.4\textwidth}
    \includegraphics[width=\linewidth]{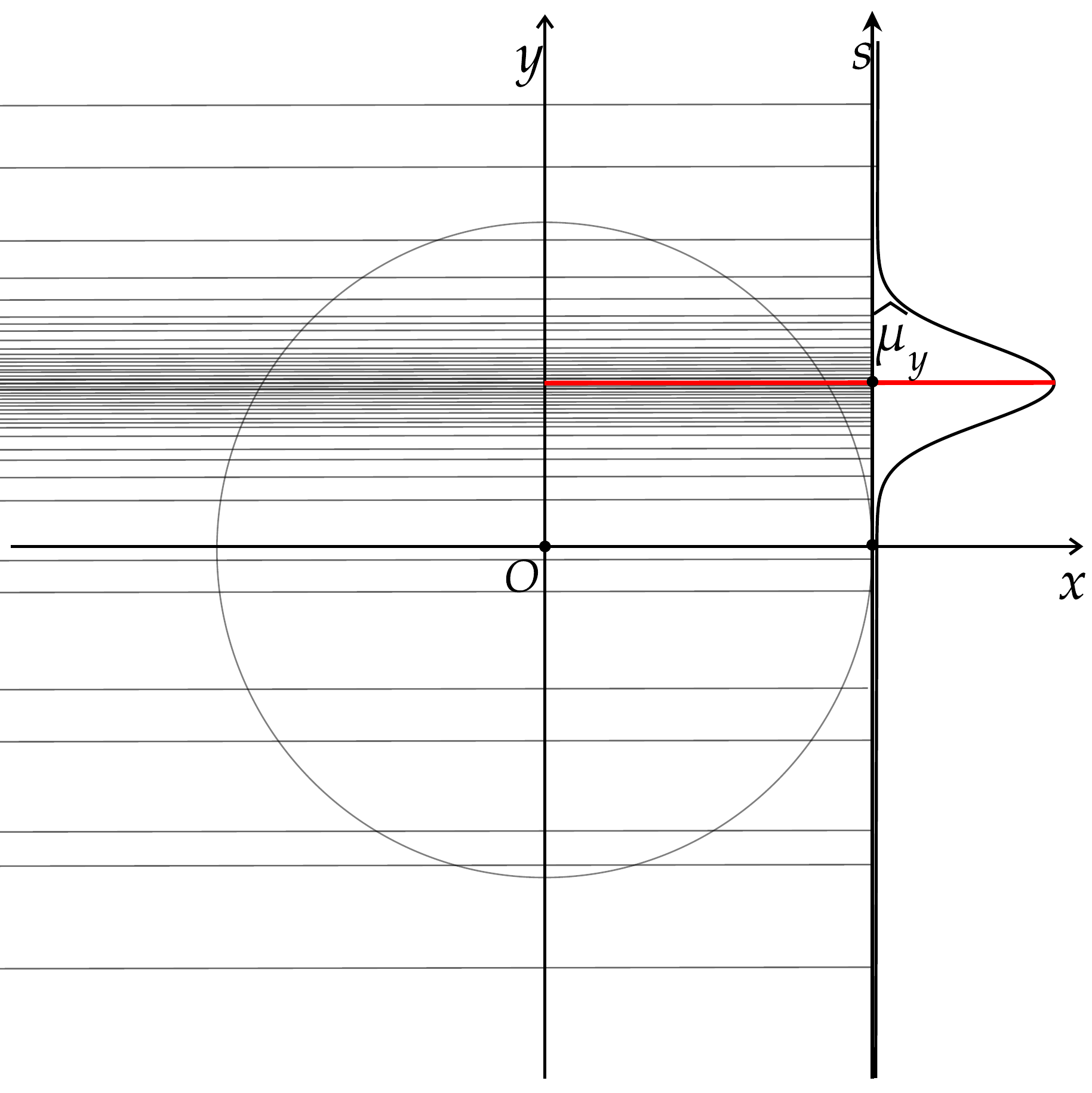}
    \caption{Estimation of $\mu_y$}
    \label{fig:mean_estim_y}
    \end{subfigure}\hspace{10pt}
    \begin{subfigure}{0.4\textwidth}
    \includegraphics[width=\linewidth]{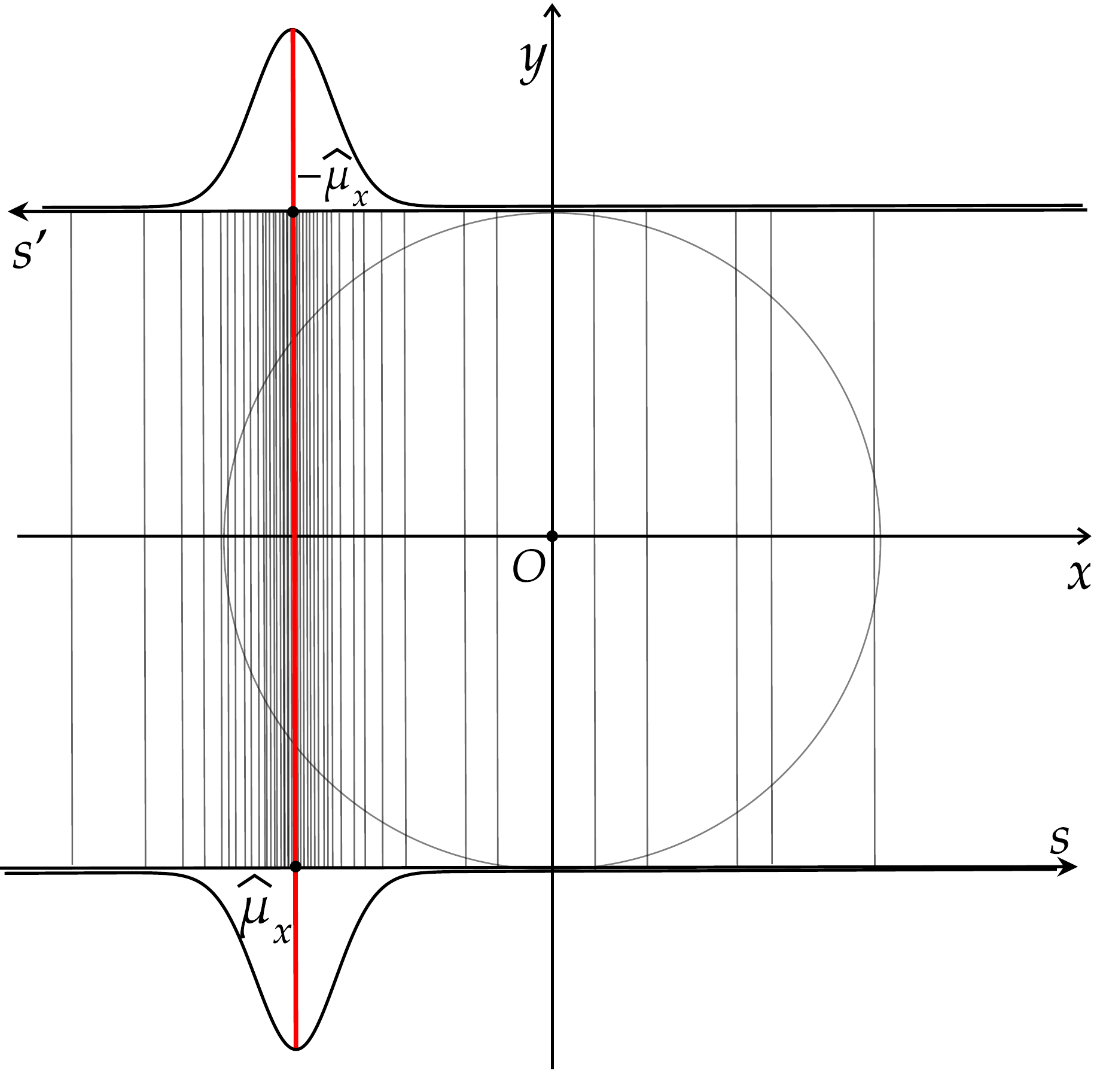}
    \caption{Estimation of $\mu_x$}
    \label{fig:mean_estim_x}
    \end{subfigure}
\caption{Estimation of mean vector} 
\label{fig:mean_estim}
\end{figure}

The point $\bm\mu_k$ (i.e. the mean vector) corresponds to a sinusoidal function in the projection domain:
\begin{equation}
m(\varphi;A,B) = A \sin(\varphi) + B \cos(\varphi),
\label{eq:mean_in_proj_unknowns}
\end{equation}
where $A\in\mathbb{R}$ and $B\in\mathbb{R}$ are unknown parameters yet to be determined. 
As depicted in Fig. \ref{fig:mean_estim_y}, we have:
\[
\mu_y^{(k)}=m(\varphi=0;A,B) = B,
\]
and Fig. \ref{fig:mean_estim_x} gives:
\[
\mu_x^{(k)}=m(\varphi = -\frac{\pi}{2};A,B) = -A.
\]
Notice that both $\varphi = -\frac{\pi}{2}$ and $\varphi = \frac{\pi}{2}$ correspond to LoR-s parallel to $y$-axis, but differ in orientation of $s$-axis. In Fig. \ref{fig:mean_estim_x}, we denoted by $s$ direction when $\varphi = -\frac{\pi}{2}$, and by $s'$ when $\varphi = \frac{\pi}{2}$. We can rewrite Eq. \ref{eq:mean_in_proj_unknowns} as
\begin{equation}
m(\varphi;\mu_x^{(k)},\mu_y^{(k)}) = -\mu_x^{(k)} \sin(\varphi) + \mu_y^{(k)} \cos(\varphi),
\label{eq:mean_in_proj}
\end{equation}
where $\bm \mu_k = \bigl( \begin{smallmatrix}
  \mu_x^{(k)}\\ \mu_y^{(k)}
\end{smallmatrix} \bigr)$ is the mean vector of $k$-th Gaussian component.

We know that the projection under an arbitrary angle of a bivariate normal distribution is a univariate normal distribution. Combining Eq. \ref{eq:mean_in_proj} and Eq. \ref{eq:uni_ls} results in a weighted least square problem: 
\begin{equation}
\begin{gathered}
L(\mu_x^{(k)},\mu_y^{(k)};s_i,\varphi_i) = \displaystyle\sum_{i=1}^N p_{ij}(m(\varphi_i;\mu_x^{(k)},\mu_y^{(k)}) - s_i)^2 = \\
\displaystyle\sum_{i=1}^N p_{ik} (-\mu_x^{(k)} \sin(\varphi_i) + \mu_y^{(k)} \cos(\varphi_i) - s_i)^2,
\end{gathered}
\label{eq:lsp_mean_in_proj}
\end{equation}
where $(s_i,\varphi_i)$ represent $i$-th LoR, $N$ is the total number of LoR-s, and weights $p_{ik}$ are membership probabilities between $i$-th LoR and $k$-th Gaussian component. The calculation of the membership probabilities is given in Section 4. 

To solve the least square problem, we set the gradient of the $L$ function to zero. We get $2\times2$ linear system $\bm M_k \bm \mu_k= \bm b_k$, where:
\begin{equation}
\begin{gathered}
\bm M_k =  \begin{pmatrix} -\displaystyle\sum_{i=1}^N p_{ik} \sin^2(\varphi_i)                 & \displaystyle\sum_{i=1}^N p_{ik} \sin(\varphi_i)\cos(\varphi_i)\\
                  -\displaystyle\sum_{i=1}^N p_{ik} \sin(\varphi_i)\cos(\varphi_i)      & \displaystyle\sum_{i=1}^N p_{ik} \cos^2(\varphi_i) 
    \end{pmatrix}\\
\bm b_k = \begin{pmatrix} \displaystyle\sum_{i=1}^N p_{ik} s_i \sin(\varphi_i)\\
                  \displaystyle\sum_{i=1}^N p_{ik} s_i\cos(\varphi_i)
    \end{pmatrix}.\\
\end{gathered}
\end{equation}
As in (\ref{eq:uni_ls_sol}), the solution is estimated mean $\bm\hat{\bm\mu_k} =   \bigl( \begin{smallmatrix}
 \hat{\mu}_x^{(k)}\\ \hat{\mu}_y^{(k)}
\end{smallmatrix} \bigr):$
\begin{equation}
\bm\hat{\bm\mu}_k = \bm M^{-1}_k \bm b_k.
\label{eq:solve_mu}
\end{equation}


\subsection{Estimation of Covariance Matrix}

Once we have the mean $\hat{\bm\mu}_k$, we need to estimate the covariance matrix $\bm\Sigma_k$. Still, we focus only on one component in the mixture, and our approach is based on the variance of the projected univariate normal distribution.

We observe a centered ($\bm\mu=\bm0$) bivariate normal distribution whose principal axes are the $x$ and $y$, without loss of generality. For any bivariate normal distribution, we simply define a new coordinate system where the new origin is the mean vector of the distribution (translation), and axes that are the eigenvectors of the covariance matrix (rotation). 

\begin{figure}[h]
\centering
    \begin{subfigure}{0.4\textwidth}
    \includegraphics[width=\linewidth]{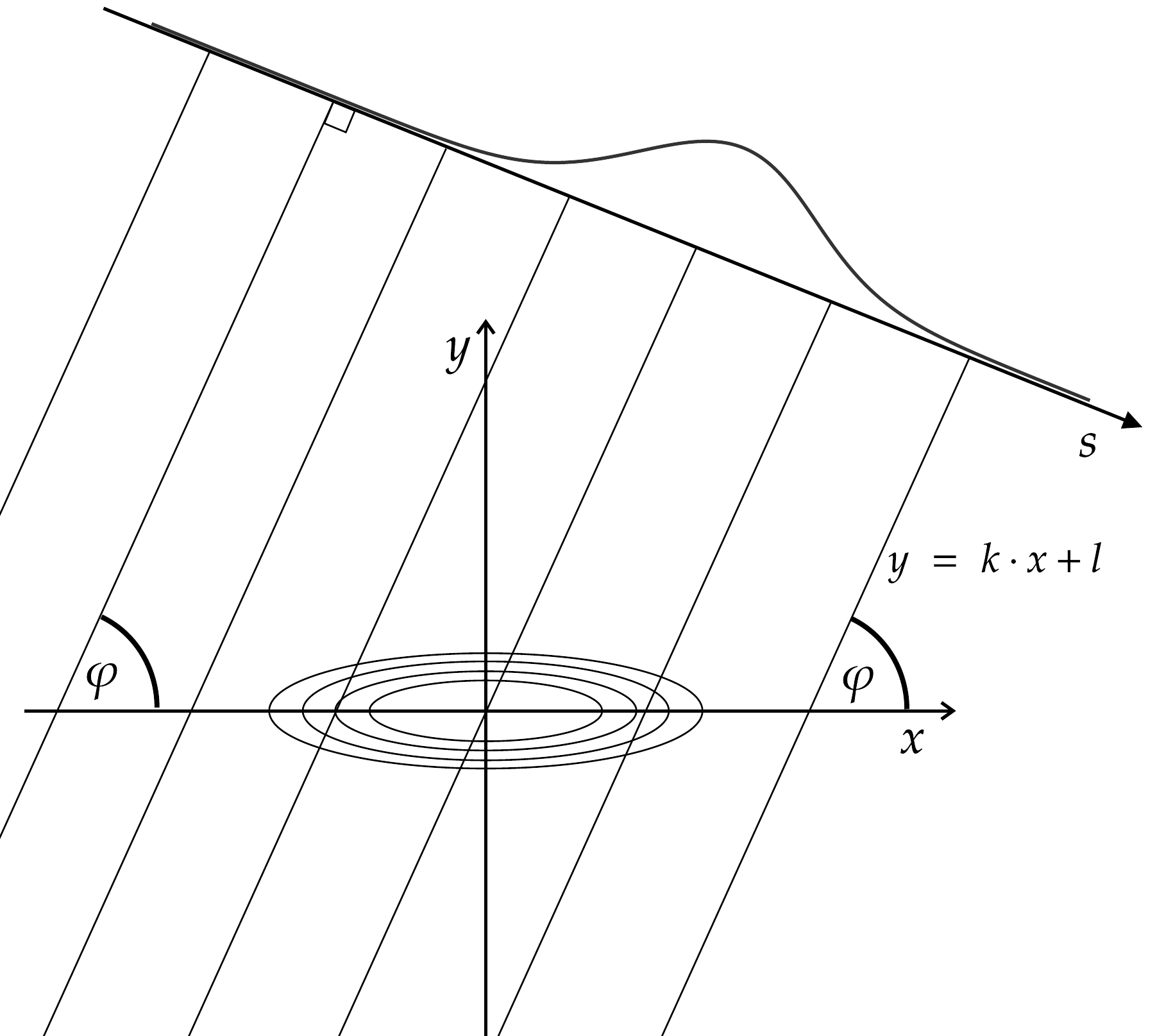}
    \caption{Line integral (projection) under angle $\varphi$}
    \label{fig:line_integral}
    \end{subfigure}\hspace{1mm}
    \begin{subfigure}{0.35\textwidth}
    \includegraphics[width=\linewidth]{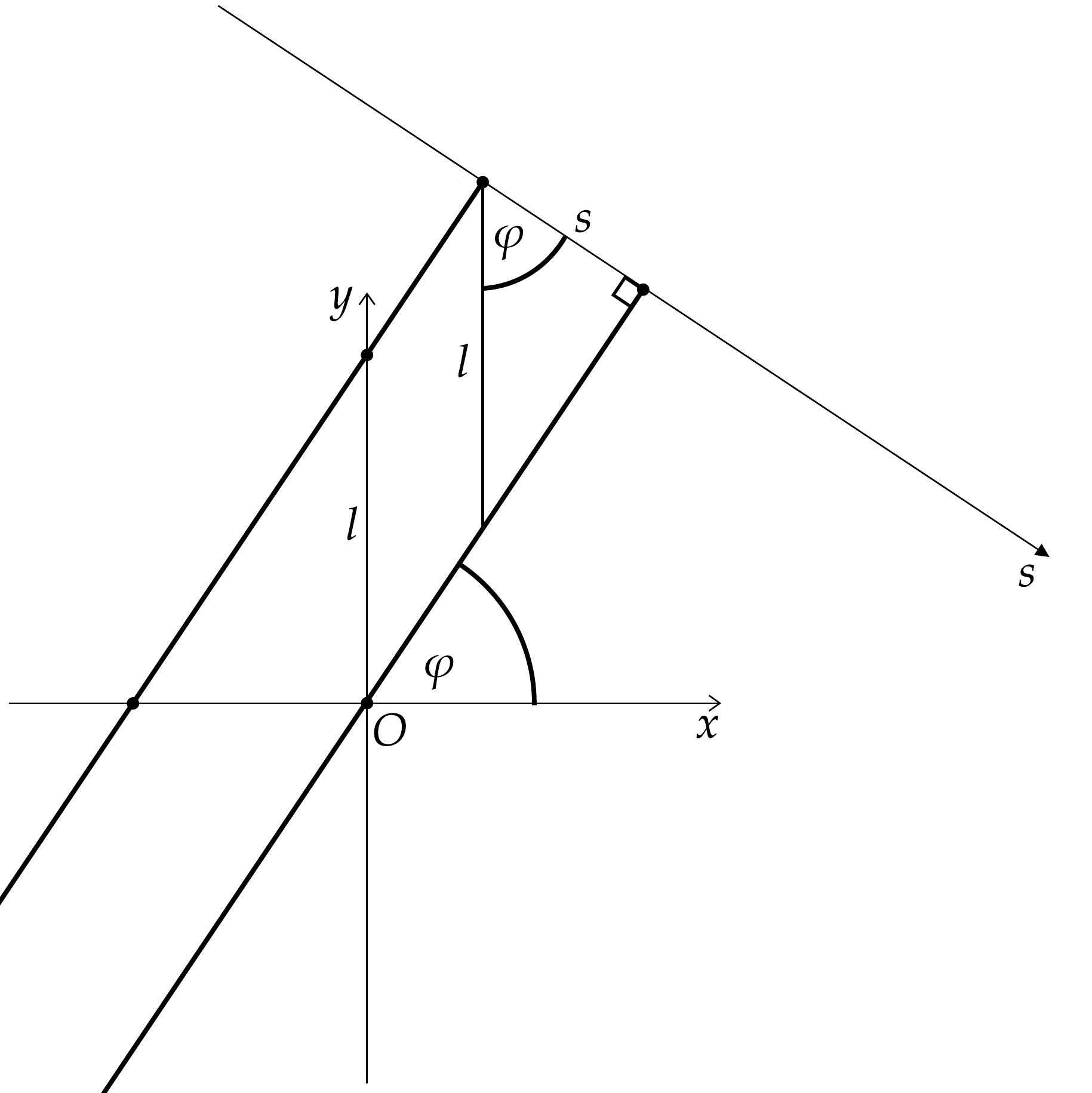}
    \caption{Relation between $y$-intercept and $s$ axis}
    \label{fig:l_and_s}
    \end{subfigure}\\
    \begin{subfigure}{0.4\textwidth}
    \includegraphics[width=\linewidth]{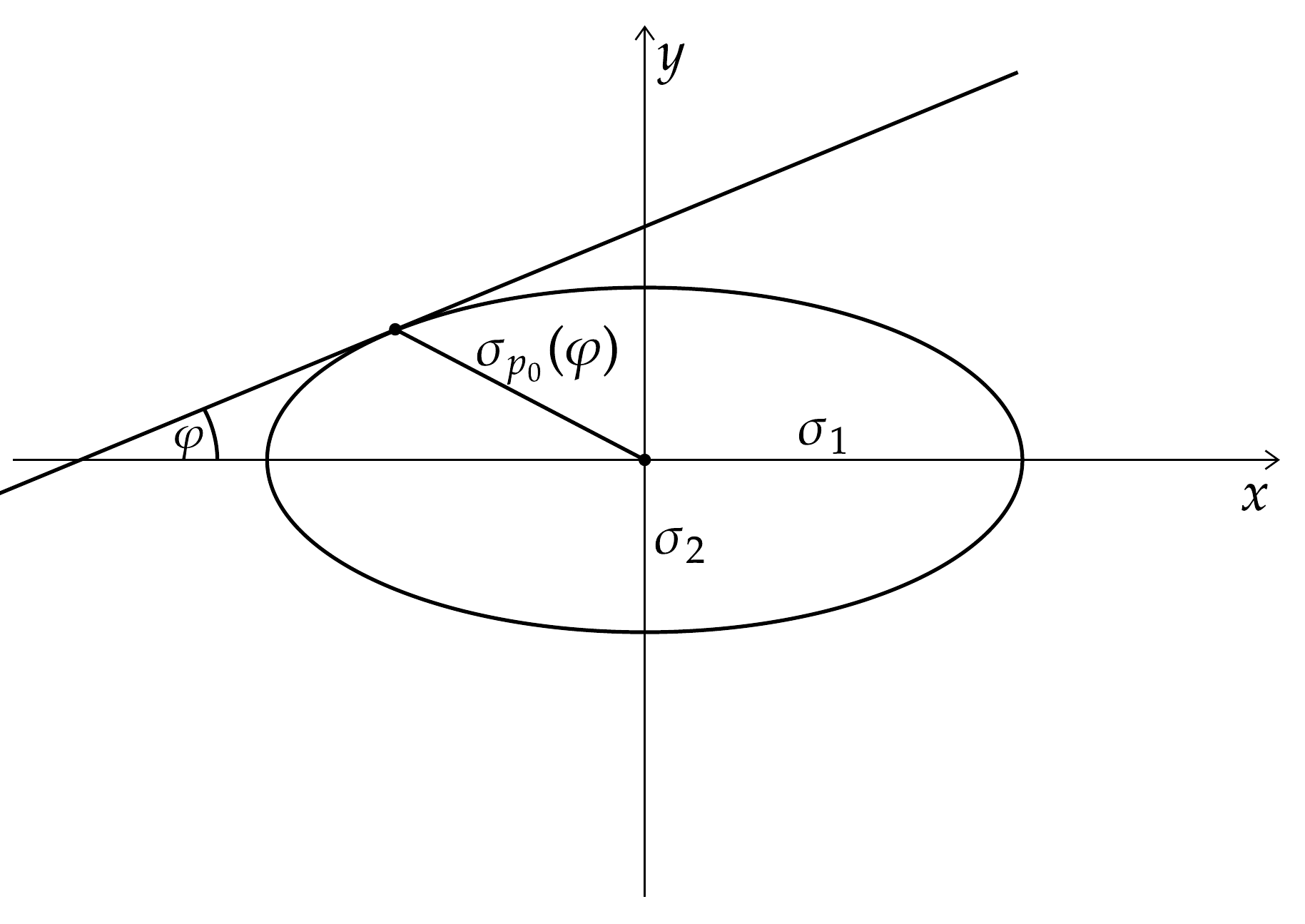}
    \caption{Ellipse, $\sigma_{p_0}(\varphi)$}
    \label{fig:ellipse_xy}
    \end{subfigure}\hspace{1mm}
    \begin{subfigure}{0.4\textwidth}
    \includegraphics[width=\linewidth]{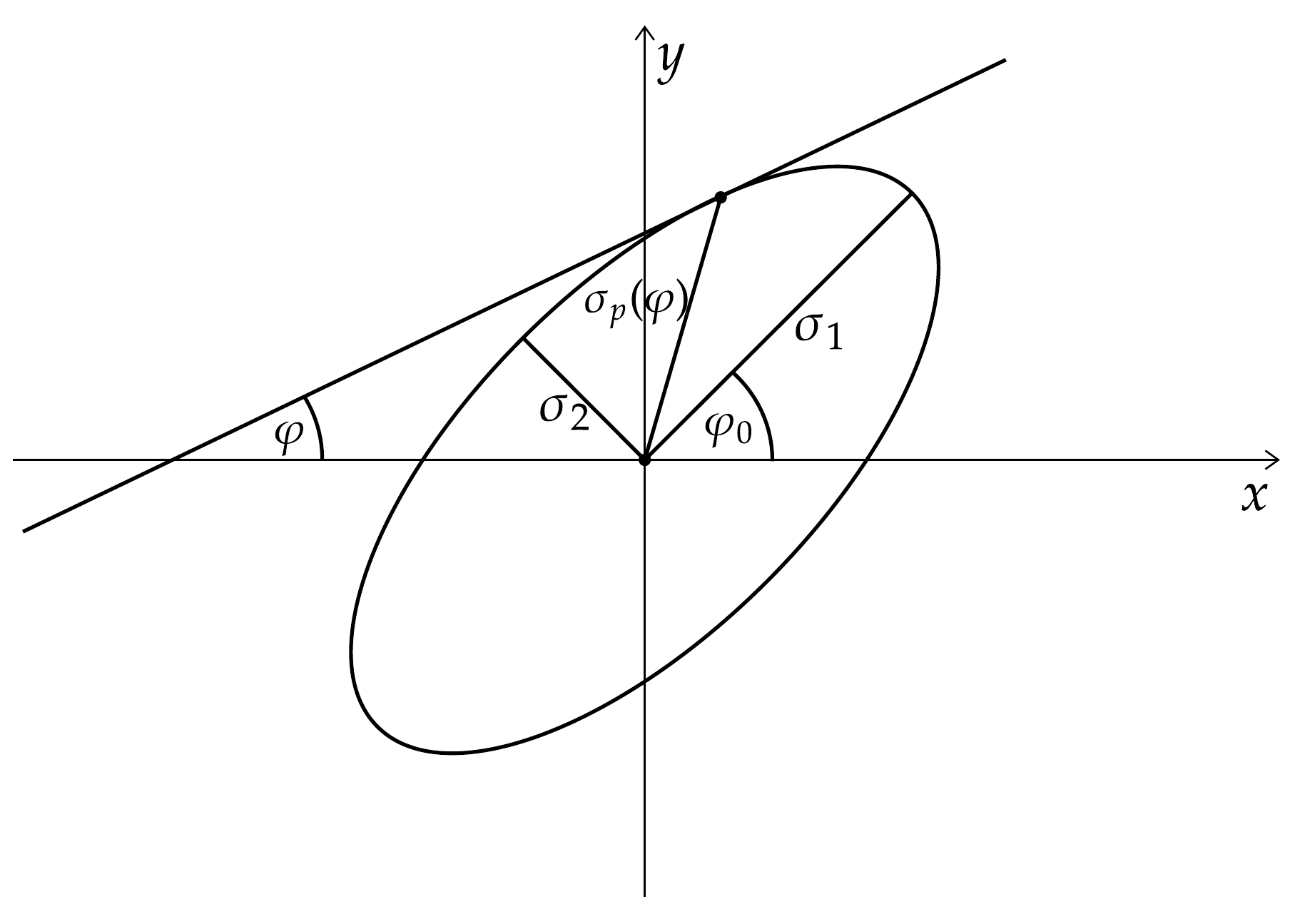}
    \caption{Rotated ellipse, $\sigma_p(\varphi)$}
    \label{fig:ellipse_rot}
    \end{subfigure}
\caption{Projection of bivariate normal distribution and elliptical illustration of its variance} 
\label{fig:integral_gauss}
\end{figure}

A centered bivariate normal distribution whose principal axes are the $x$ and $y$ can be expressed as:
\[
f_{Gc}(x,y) = \frac{1}{2\pi\sigma_1\sigma_2}\exp(-(\frac{x^2}{2\sigma_1^2}+\frac{y^2}{2\sigma_2^2})).
\]
We want to find the distribution under some angle $\varphi$ as seen in Fig. \ref{fig:line_integral}. We must calculate integral:
\begin{equation}
f_{p_{0_\varphi}}(s)=\int_{\gamma_{\varphi}}f_{Gc}(x,y)d\gamma_{\varphi},
\label{eq:normal_integrate}
\end{equation}
where $\gamma_{\varphi}(t) = (t,kt+l)$, $k=\tan(\varphi)$ and $l=\frac{s}{\cos(\varphi)}$. Mapping  $\gamma_{\varphi}$ parameterizes the set of all lines under the angle $\varphi$. Fig. \ref{fig:l_and_s} explains the connection between y-intercept ($l$) and $s$-axis. After integration we get
\begin{equation}
f_{p_{0_\varphi}}(s) = \frac{1}{\sqrt{2\pi}\sigma_{p_0}(\varphi)}\exp(-\frac{s^2}{2\sigma^2_{p_0}(\varphi)}),
\end{equation} 
where $\sigma_{p_0}^2(\varphi) = {\sigma_1^2\sin^2(\varphi)+\sigma_2^2\cos^2(\varphi)}$. A detailed calculation can be found in \ref{AppA}. Function $\sigma_{p_0}(\varphi)$ can be interpreted as the distance from the origin to one of the tangent lines on the ellipse $\frac{x^2}{\sigma_1^2} + \frac{y^2}{\sigma_2^2} = 1$, under angle $\varphi$, as shown in Fig. \ref{fig:ellipse_xy}. If we substitute $\varphi$ with $\varphi - \varphi_0$ in $\sigma_{p_0}(\varphi)$, we get more general expression for the variance of some rotated Gaussian (i.e. ellipse). It corresponds to the rotation of the entire coordinate system by angle $\varphi_0$.

The expression is now:
\[
\sigma^2_p(\varphi) = {\sigma_1^2\sin^2(\varphi-\varphi_0) + \sigma_2^2\cos^2(\varphi-\varphi_0)}, 
\]
as shown in Fig. \ref{fig:ellipse_rot}. The probability density function of the projected and centered ($\bm\mu=\bm 0$) bivariate normal distribution under angle $\varphi_0$ is:
\begin{equation}
f_{p_\varphi}(s;\sigma_1,\sigma_2,\varphi_0) = \frac{1}{\sqrt{2\pi}\sigma_{p}(\varphi)}\exp(-\frac{s^2}{2\sigma^2_{p}(\varphi)}).
\label{eq:pdf_1d}
\end{equation}

\begin{figure}[th]
\centering
    \begin{subfigure}{0.4\textwidth}
    \includegraphics[width=\linewidth]{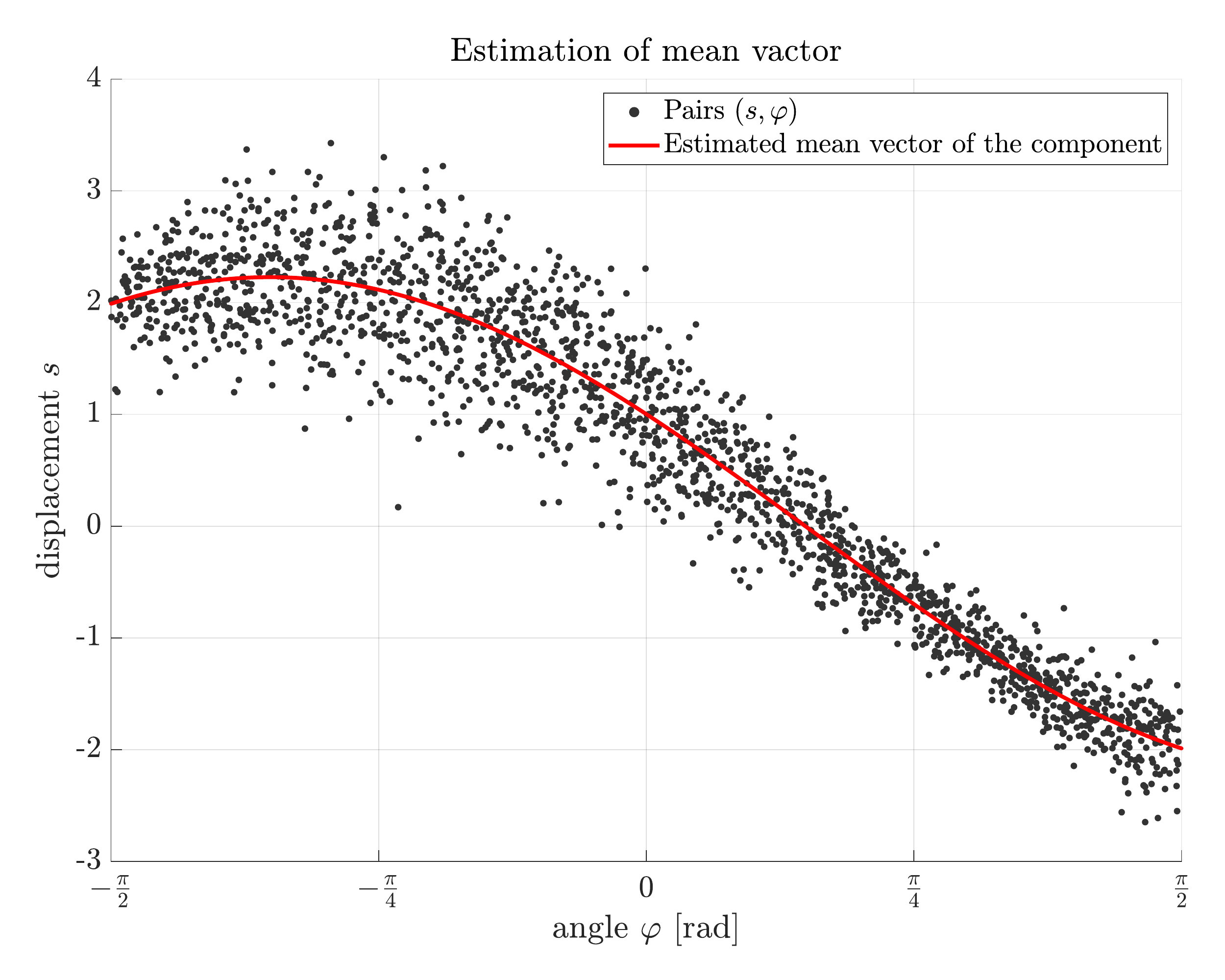}
    \caption{Estimation of mean value in the projection domain}
    \label{fig:onecomp_mean}
    \end{subfigure}\hspace{1mm}
    \begin{subfigure}{0.4\textwidth}
    \includegraphics[width=\linewidth]{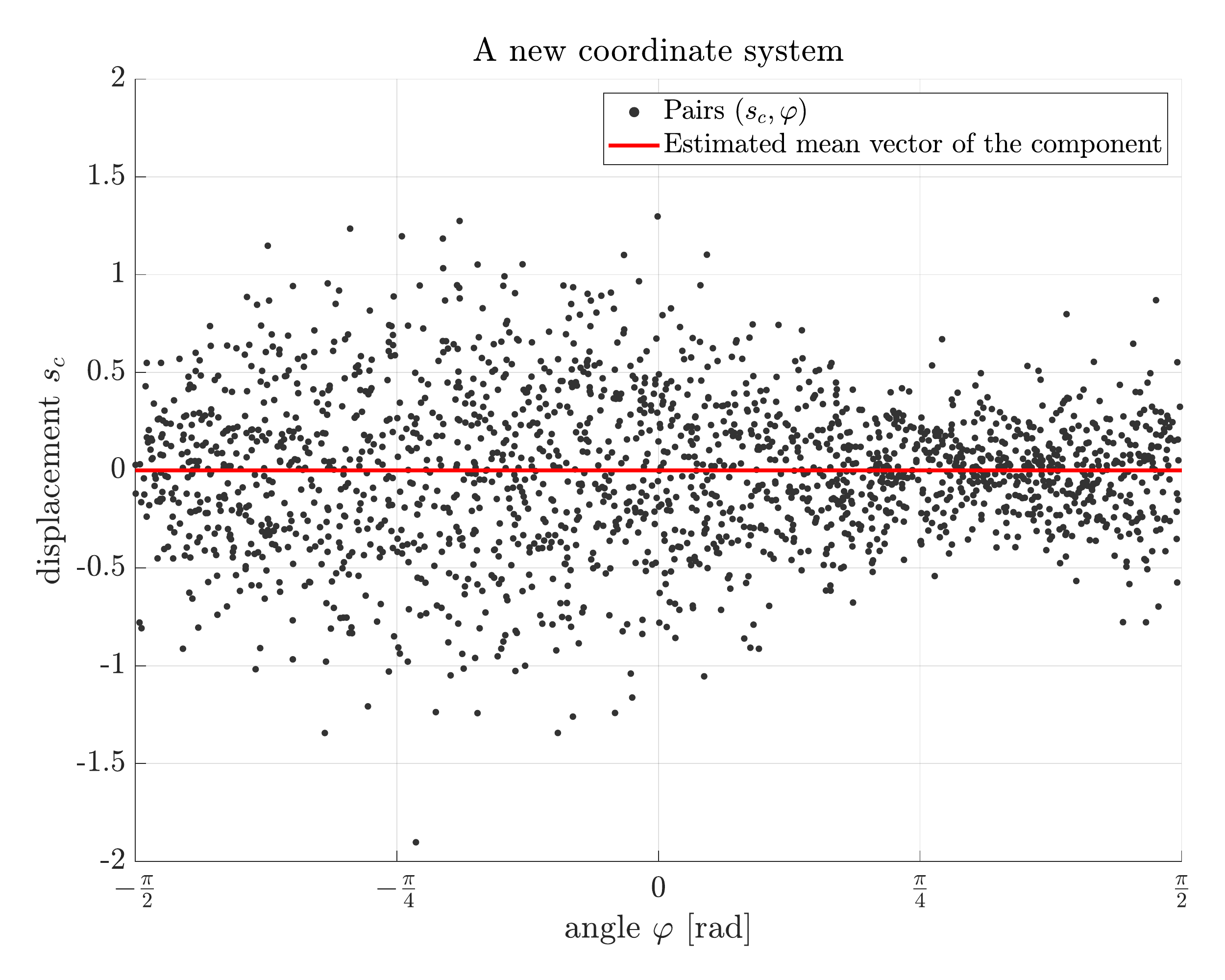}
    \caption{Pairs $(s_c,\varphi)$ in a new coordinate system}
    \label{fig:centered_mean}
    \end{subfigure}\\
\caption{Estimation of mean value and LoR-s in the original (left) and centered (right) coordinate systems} 
\label{fig:means}
\end{figure}

For $k$-th component, additionally we need to translate the coordinate system by mean vector $\bm \mu_k$. In the projection space it is:
\begin{equation}
(s_c^{(k)},\varphi) = (s - m(\varphi;\mu_x^{(k)},\mu_y^{(k)}), \varphi). 
\label{eq:centered_pairs}
\end{equation}
A point $(s_c^{(k)},\varphi)$ is a LoR in the centered coordinate system. In Fig. \ref{fig:means}, we can see points $(s,\varphi)$ and the mean before (Fig. \ref{fig:onecomp_mean}) and after translation (Fig. \ref{fig:centered_mean}). 

Consider $N$ LoR-s $(s_{c_i}^{(k)}, \varphi_i)$. Samples $\varphi_i$ follow a uniform distribution on the interval $[-\frac{\pi}{2}, \frac{\pi}{2}]$, due to the nature of electron-positron annihilation. Interestingly, $s_{c_i}^{(k)}$ follow the distribution whose probability density function is:
\begin{equation}
f_{s_c^{(k)}}(s_c;\sigma_1^{(k)},\sigma_2^{(k)},\varphi_0^{(k)})=\displaystyle \frac{1}{\pi}\int\displaylimits_{-\frac{\pi}{2}}^{\frac{\pi}{2}} \frac{1}{\sqrt{2\pi}\sigma_p(\varphi)} e^{-\frac{s_c^2}{2\sigma^2_p(\varphi)}}d\varphi.
\label{eq:profil_pdf}
\end{equation}
This expression comes from the fact that for fixed angle $\varphi$ we have a probability density function $f_{p_\varphi}(s_c;\sigma_1^{(k)},\sigma_2^{(k)},\varphi_0^{(k)})$. By averaging all contributions for all angles $\varphi$, we get the expression (\ref{eq:profil_pdf}), which is the marginal distribution of $s_c^{(k)}$. 

Similarly to the problem from Eq. \ref{eq:uni_ls}, let $(x_i)_{i=1}^N$ be the samples that follow distribution $f_{p_\varphi}$ from Eq. \ref{eq:pdf_1d} for some fixed angle $\varphi$. The least square problem:
\begin{equation}
L_{sq}(d;x_i) = \displaystyle\sum_{n=1}^N (d-x_i^2)^2
\label{eq:sq_ls}
\end{equation}
has the solution
\begin{equation}
d_{LS} = \frac{1}{N} \displaystyle\sum_{n=1}^N x_i^2 \approx E[X^2] = \sigma^2_p(\varphi),
\label{eq:sq_ls_sol}
\end{equation}
where $\sigma^2_p(\varphi)$ is the variance of the distribution with the probability density function $f_{p_\varphi}$. Therefore, we can state the weighted least square problem:
\begin{equation}
\begin{gathered}
L_{sq}(\varphi_0^{(k)},\sigma_1^{(k)},\sigma_2^{(k)}) = \\ \displaystyle\sum\displaylimits_{i=1}^N p_{ik}\cdot [(\sigma_1^{(k)})^2\sin^2(\varphi_i-\varphi_0^{(k)}) + (\sigma_2^{(k)})^2\cos^2(\varphi_i-\varphi_0^{(k)}) -(s_{c_i}^{(k)})^2]^2.
\label{eq:var_lsp}
\end{gathered}
\end{equation}
Calculating $\nabla L_{sq}(\varphi_0^{(k)},\sigma_1^{(k)},\sigma_2^{(k)}) = \bm 0$ leads to a nonlinear system of equations with no closed-form solution. 
To solve this system, we need to find the roots of a polynomial with an order greater than four, which can only be solved numerically. 

Therefore, in this paper, we propose a different approach. If we assume that $\sigma_1^{(k)}$ and $\sigma_2^{(k)}$ are known, and focus only on minimizing (\ref{eq:var_lsp}) with respect to rotation $\varphi_0^{(k)}$, we get a simple polynomial system with a closed-form solution. Similarly, once we know $\varphi_0^{(k)}$, and want to find $\sigma_1^{(k)}$ and $\sigma_2^{(k)}$ that solve the problem stated by the Eq. \ref{eq:var_lsp}, we get a linear $2\times2$ system of equations.

Let us put the moments of the probability density function $f_{s_c}$ in relation with the variances $\sigma_1^{(k)}$ and $\sigma_2^{(k)}$. Notice that odd moments are zero since $f_{s_c}$ is an even function. We calculate the second and the fourth moment. The second moment is
\begin{equation}
\begin{gathered}
E[S_c^2] =\frac{1}{\pi}\displaystyle\int\displaylimits_{-\infty}^{+\infty} \displaystyle\int\displaylimits_{-\frac{\pi}{2}}^{\frac{\pi}{2}} s_c^2 f_{s_c}(s_c;\sigma_1^{(k)},\sigma_2^{(k)},\varphi_0^{(k)})d\varphi ds_c=
 \frac{1}{\pi}\displaystyle\int\displaylimits_{-\frac{\pi}{2}}^{\frac{\pi}{2}} \sigma_p^2(\varphi)=\\
 \frac{1}{\pi}\displaystyle\int\displaylimits_{-\frac{\pi}{2}}^{\frac{\pi}{2}} [(\sigma_1^{(k)})^2\cos^2(\varphi-\varphi_0^{(k)})  + (\sigma_2^{(k)})^2\sin^2(\varphi-\varphi_0^{(k)})]d\varphi = 
 \frac{(\sigma_1^{(k)})^2}{2} + \frac{(\sigma_2^{(k)})^2}{2},
\end{gathered}
\label{eq:second_moment}
\end{equation}
while the fourth moment is
\begin{equation}
\begin{gathered}
E[S_c^4] =\frac{1}{\pi}\displaystyle\int\displaylimits_{-\infty}^{+\infty} \displaystyle\int\displaylimits_{-\frac{\pi}{2}}^{\frac{\pi}{2}} s_c^4 f_{s_c}(s_c;\sigma_1^{(k)},\sigma_2^{(k)},\varphi_0^{(k)})d\varphi ds_c=
  \frac{1}{\pi}\displaystyle\int\displaylimits_{-\frac{\pi}{2}}^{\frac{\pi}{2}} 3\sigma_p^4(\varphi)=\\
 \frac{3}{\pi}\displaystyle\int\displaylimits_{-\frac{\pi}{2}}^{\frac{\pi}{2}} [(\sigma_1^{(k)})^2\cos^2(\varphi-\varphi_0^{(k)})+  (\sigma_2^{(k)})^2\sin^2(\varphi-\varphi_0^{(k)})]^2d\varphi =\\
 \frac{9(\sigma_1^{(k)})^4}{8} +  \frac{3(\sigma_1^{(k)})^2(\sigma_2^{(k)})^2}{4} +  \frac{9(\sigma_2^{(k)})^4}{8}.
\end{gathered}
\label{eq:forth_moment}
\end{equation}
But, second and fourth moments can be estimated from:
\begin{equation}
\begin{gathered}
M_{2w}^{(k)} = \displaystyle\frac{\sum_{i=1}^N p_{ik} (s_{c_i}^{(k)})^2}{\sum_{n=1}^N p_{ik}}\\
M_{4w}^{(k)} = \displaystyle\frac{\sum_{i=1}^N p_{ik} (s_{c_i}^{(k)})^4}{\sum_{n=1}^N p_{ik}}
\end{gathered},
\label{eq:sample_moments}
\end{equation}
where $s_{c_i}^{(k)}$ are samples as described in Eq. \ref{eq:centered_pairs} and $p_ {ik}$ is the probability that $i$-th LoR belongs to the $k$-th component of the mixture.

From Eq. \ref{eq:second_moment}, \ref{eq:forth_moment}, and \ref{eq:sample_moments} we can estimate $(\sigma_1^{(k)})^2$ and $(\sigma_2^{(k)})^2$. By taking the square root, we get $\sigma_1^{(k)}$ and $\sigma_2^{(k)}$. The solution of system:
\begin{equation}
\begin{gathered}
\frac{(\hat{\sigma}_1^{(k)})^2}{2} + \frac{(\hat{\sigma}_2^{(k)})^2}{2} = M_{2w}^{(k)}\\
\frac{9(\hat{\sigma}_1^{(k)})^4}{8} +  \frac{3(\hat{\sigma}_1^{(k)})^2(\hat{\sigma}_2^{(k)})^2}{4} +  \frac{9(\hat{\sigma}_2^{(k)})^4}{8} = M_{4w}^{(k)} 
\end{gathered}
\label{eq:moment_system}
\end{equation}
gives estimates $(\hat{\sigma}_1^{(k)})^2$ and $(\hat{\sigma}_2^{(k)})^2$. By expressing $(\hat{\sigma}_2^{(k)})^2$ from the first equation and inserting it into the second one, we get a quadratic equation in variable $(\hat{\sigma}_1^{(k)})^2$:
\begin{equation}
\begin{gathered}
3 (\hat{\sigma}_1^{(k)})^4 - 6M_{2w} (\hat{\sigma}_1^{(k)})^2 + 9M_{2w}^{(k)} - 2M_{4w}^{(k)} = 0
\end{gathered}.
\label{eq:moment_system_quad}
\end{equation}
Notice that the system in Eq. \ref{eq:moment_system} is symmetric, so we can choose $(\hat{\sigma}_1^{(k)})^2 \geq (\hat{\sigma}_2^{(k)})^2$. The solution of the system (\ref{eq:moment_system}) is given by Eq. \ref{eq:moment_system_quad}, and the final solution is:
\begin{equation}
\begin{gathered}
(\hat{\sigma}_1^{(k)})^2  = M_{2w}^{(k)} + \sqrt{2}\cdot\sqrt{\frac{M_{4w}^{(k)}}{3} - (M_{2w}^{(k)})^2},\\
(\hat{\sigma}_2^{(k)})^2  = M_{2w}^{(k)} - \sqrt{2}\cdot\sqrt{\frac{M_{4w}^{(k)}}{3} - (M_{2w}^{(k)})^2}.
\end{gathered}
\label{eq:moment_sol}
\end{equation}
Now, we can calculate $\varphi_0^{(k)}$ by solving the weighed least square problem in (\ref{eq:var_lsp}). By using trigonometric power-reduction formulae we can rewrite Eq. \ref{eq:var_lsp} as:
\begin{equation}
\begin{gathered}
L_{sq}(\varphi_0^{(k)},\sigma_1^{(k)},\sigma_2^{(k)})) =\\ \displaystyle\sum\displaylimits_{i=1}^N p_{ik}\cdot \bigg(\frac{(\sigma_2^{(k)})^2-(\sigma_2^{(k)})^2}{2}\cos(2\varphi_0^{(k)} - 2\varphi_i)  + \frac{(\sigma_1^{(k)})^2+(\sigma_2^{(k)})^2}{2} - (s_{c_i}^{(k)})^2\bigg)^2.
\end{gathered}
\label{eq:rewrite_var_lsp}
\end{equation}
We set the partial derivative with respect to $\varphi_0^{(k)}$ to zero, i.e. $\frac{\partial L_{sq}}{\partial\varphi_0^{(k)}}(\varphi_0^{(k)}) = 0$:
\begin{equation}
\begin{gathered}
\displaystyle\sum\displaylimits_{i=1}^N (M_{ik}\cos(\alpha_0^{(k)} - \alpha_i) + N_{ik})\sin(\alpha_0^{(k)} - \alpha_i) = 0,
\end{gathered}
\label{eq:partial_phi0}
\end{equation}
where 
\begin{equation}
\begin{gathered}
M_{ik} = p_{ik}((\sigma_2^{(k)})^2-(\sigma_2^{(k)})^2),\\
N_{ik} = p_{ik}((\sigma_1^{(k)})^2+(\sigma_2^{(k)})^2 - 2(s_{c_i}^{(k)})^2), \\
\alpha_0^{(k)} = 2\varphi_0^{(k)}, \text{and}\\
\alpha_i = 2\varphi_i.
\end{gathered}
\end{equation}
The equation can be written in a more useful form by using angle difference identities:
\begin{equation}
\begin{gathered}
A_{s2}\sin^2(\alpha_0^{(k)}) + A_{c2}\cos^2(\alpha_0^{(k)}) + A_{sc}\sin(\alpha_0^{(k)})\cos(\alpha_0^{(k)})
+\\ A_{s}\sin(\alpha_0^{(k)}) +  A_{c}\cos(\alpha_0^{(k)}) = 0,
\end{gathered}
\label{eq:Simple_form_a}
\end{equation}
where
\begin{equation}
\begin{gathered}
A_{s2} = \displaystyle\sum\displaylimits_{i=1}^N M_{ik} \sin(\alpha_i)\cos(\alpha_i), \\
A_{c2} = -\displaystyle\sum\displaylimits_{i=1}^N M_{ik} \sin(\alpha_i)\cos(\alpha_i) = - A_{s2}, \\
A_{sc} = \displaystyle\sum\displaylimits_{i=1}^N M_{ik}\cos^2(\alpha_i) - \displaystyle\sum\displaylimits_{i=1}^N M_{ik}\sin^2(\alpha_i), \\
A_{s} = \displaystyle\sum\displaylimits_{i=1}^N N_{ik}\cos(\alpha_i), \\
A_{c} = - \displaystyle\sum\displaylimits_{i=1}^N N_{ik}\sin(\alpha_i).\\
\end{gathered}
\label{eq:Simple_form}
\end{equation}
Introducing substitution $x=\cos(\alpha_0^{(k)})$ and $y=\sin(\alpha_0^{(k)})$, we get a system of two polynomials equations:
\begin{equation}
\begin{cases}
\begin{gathered}
x^2+y^2 = 1,\\
A_{s2}y^2 - A_{s2}x^2 + A_{sc}xy + A_{s}y +  A_{c}x = 0.
\end{gathered}
\end{cases}
\label{eq:system_phi}
\end{equation}
Now, we substitute $y = \pm \sqrt{1-x^2}$ in the second equation and calculate:
\begin{equation}
\begin{gathered}
\pm(A_{sc}x + A_s)\sqrt{1-x^2} = 2A_{s2}x^2 - A_cx + A_{s2} \Big/^2 \Rightarrow\\
(4A_{s2}^2+A_{sc}^2)x^4 + (2A_{sc}A_s-4A_{s2}A_c)x^3+
(A_c^2+A_s^2-A_{sc}^2-4A_{s2}^2)x^2+\\(2A_{s2}A_c-2A_{sc}A_s)x+
A_{s2}^2-A_s^2 = 0
\end{gathered}
\label{eq:system_phi_poly}
\end{equation}
Therefore, we get a quartic equation that has an algebraic (close-form) solution as stated by Abel-Ruffini theorem. More details on how to find roots of a polynomial of the forth order analytically can be found in \cite{Fathi2013}. There are eight pairs of solutions. Four of them do not satisfy the second equation in (\ref{eq:system_phi}), leaving us with only four solutions. Two of the remaining candidates are complex conjugates. Thus, only two real solutions remain - global minimum and global maximum. We pick the solution $\hat{\varphi}_0^{(k)}=\frac{\hat{\alpha}_0^{(k)}}{2} = \frac{1}{2} \arctan(\frac{y}{x})$ with the minimum value of $L_{sq}(\hat{\varphi}_0^{(k)},\hat{\sigma}_1^{(k)},\hat{\sigma}_2^{(k)})$, which is the solution of the least square problem stated in Eq. \ref{eq:rewrite_var_lsp}.

Pay attention that now we have all data needed for construction of the covariance matrix, as shown in Eq. \ref{eq:eigo}.

The last step is to reevaluate $\sigma_1^{(k)}$ and $\sigma_2^{(k)}$ from the Eq. \ref{eq:var_lsp} when angle $\varphi_0^{(k)}$ is known. This step is optional, but useful, especially when the components significantly overlap. To solve the least square problem from Eq. \ref{eq:var_lsp}, we set partial derivatives with respect to $(\sigma_1^{(k)})^2$ and $(\sigma_2^{(k)})^2$ to zero ( $\frac{\partial L_{sq}}{\partial(\sigma_1^{(k)})^2}=\frac{\partial L_{sq}}{\partial(\sigma_2^{(k)})^2}=0$ ). 
We get a linear $2\times 2$ system: $ \underbrace{\begin{pmatrix} M_{11}              & M_{12}\\
                  M_{21}    & M_{22}
    \end{pmatrix}}_{\bm M_k} \bm\sigma^2_k = \bm b_k$, where
\begin{equation}
\begin{gathered}
M_{11}= \displaystyle\sum_{i=1}^N p_{ik} \sin^4(\varphi_0^{(k)}-\varphi_i),\\
M_{12}=M_{21} = \displaystyle\sum_{i=1}^N p_{ik} \sin^2(\varphi_0^{(k)}-\varphi_i)\cos^2(\varphi_0^{(k)}-\varphi_i),\\
M_{22} =  \displaystyle\sum_{i=1}^N p_{ik} \cos^4(\varphi_0^{(k)}-\varphi_i),\\
\bm b_k = \begin{pmatrix} \displaystyle\sum_{i=1}^N p_{ik} (s_{c_i}^{(k)})^2 \sin^2(\varphi_0^{(k)}-\varphi_i)\\
                  \displaystyle\sum_{i=1}^N p_{ik} (s_{c_i}^{(k)})^2 \cos^2(\varphi_0^{(k)}-\varphi_i)^2
    \end{pmatrix},\\
\bm \sigma^2_k = \begin{pmatrix} (\sigma_1^{(k)})^2\\
                  (\sigma_2^{(k)})^2
    \end{pmatrix}.\\
\end{gathered}
\label{eq:system_sigmas}
\end{equation}
The solution of the above system is $\hat{\bm\sigma}^2_k = \bm M_k^{-1} \bm b_k$.

\begin{figure}[th]
\centering
 \includegraphics[width=0.55\linewidth]{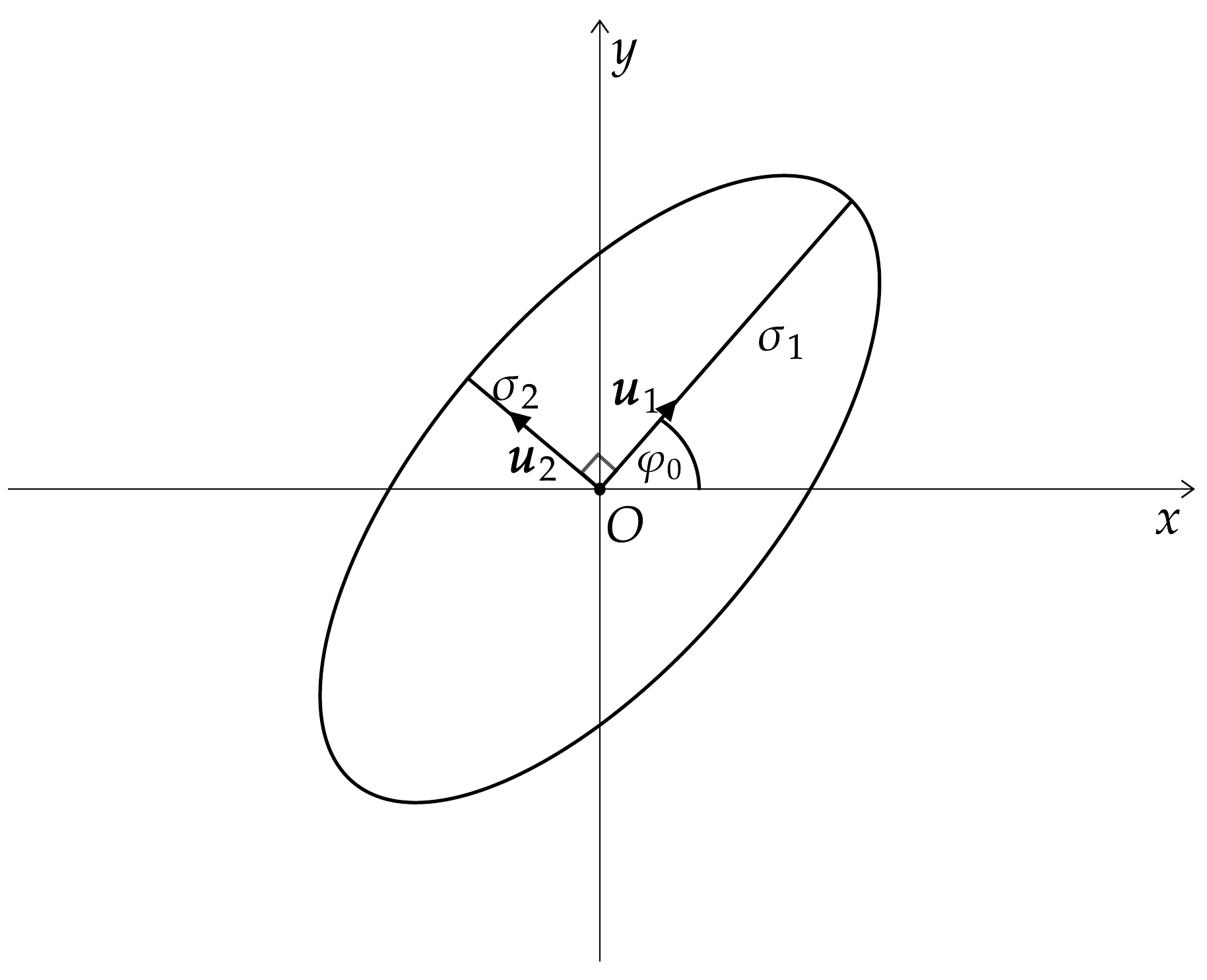}
\caption{Interpretation of eigenvectors and eigenvalues of covariance matrix} 
\label{fig:eig}
\end{figure}

Finally, from known estimates $\hat{\sigma_1}^{(k)}$, $\hat{\sigma}_2^{(k)}$ and $\hat{\varphi}_0^{(k)}$ we can construct covariance matrix $\bm\Sigma_k$. The covariance matrix can be decomposed as:
\begin{equation}
\bm \Sigma_k = \bm U_k \bm D_k \bm U_k^{T}
\label{eq:eigo}
\end{equation}
where $\bm D_k = diag((\hat{\sigma}_1^{(k)})^2,(\hat{\sigma}_2^{(k)})^2)$ is a diagonal matrix and $\bm U_k = [\bm u_1 \;\; \bm u_2]$ is a unitary (rotation) matrix. Corresponding eigenvectors $\bm u_1 $ and $\bm u_2$ are unit vectors in direction of the principal axes. Eigenvectors can be expressed in terms of angle $\varphi_0^{(k)}$ as
\begin{equation}
\begin{gathered}
\bm u_1 = \begin{pmatrix}   \cos(\hat{\varphi}_0^{(k)} )\\
                            \sin(\hat{\varphi}_0^{(k)})       
    \end{pmatrix},  \\
\bm u_2 = \begin{pmatrix}   \cos(\hat{\varphi}_0^{(k)} + \frac{\pi}{2})\\
                            \sin(\hat{\varphi}_0^{(k)} + \frac{\pi}{2})    
    \end{pmatrix}.
\end{gathered}
\end{equation}
In Fig. \ref{fig:eig}, we can see eigenvectors and the corresponding square root of eigenvalues $\hat{\sigma_1}^{(k)}$, $\hat{\sigma}_2^{(k)}$.
\section{Proposed Algorithm}
At the beginning, we arbitrarily initialize mean vectors, and randomly assign LoR-s to a subset of Gaussian components. We remind the reader that the number of components $K$ is known in advance. The proposed algorithm consists of two steps. Both steps are iterative, and the first step provides for a rough estimate of the unknown parameters. In each iteration of the second step we recalculate the mean vectors, the covariance matrices, as well as the membership probabilities. 

\subsection{First step - initialization}
Initially, we assign lines of response randomly to Gaussian components. Each component should have a roughly equal number of associated LoR-s $\approx\frac{N}{K}$. In this step, we do the hard (or modal) classification of LoR-s. That means that each LoR belongs to only one of the components, i.e. the membership probabilities are binary, and the number of LoR-s of $k$-th component is 
\begin{equation}
\hat{L}_k = \sum_{i=1}^{N} p_{ik}.
\label{eq:estim_num_of_lines}
\end{equation}

We calculate the mean vectors for each Gaussian component as described by Eq. \ref{eq:lsp_mean_in_proj} and \ref{eq:solve_mu}. Then, we reassign the LoR-s by pairing them to the least distanced mean vectors. Notice that the membership probabilities are still binary. We repeat the iterative procedure (mean, reassignment), until the mean vectors changes are sufficiently small.

From known probabilities $p_{ik}$ (zeros and ones) and mean vectors for each component in the mixture, we calculate the initial estimate of the covariance matrices. Thanks to the analytical results presented in the previous Section, the calculation is straightforward. We estimate $\sigma_1^{(k)}$ and $\sigma_2^{(k)}$ from moments for each component separately as described in Eq. \ref{eq:sample_moments}, \ref{eq:moment_system} and \ref{eq:moment_sol}. Afterward, we calculate $\hat{\varphi}_0^{(k)}$ as discussed in the previous Section. Then we reevaluate $\hat{\sigma}_1^{(k)}$ and $\hat{\sigma}_2^{(k)}$ as stated in Eq. \ref{eq:system_sigmas} and update $\hat{\varphi}_0^{(k)}$ with new $\hat{\sigma}_1^{(k)}$ and $\hat{\sigma}_2^{(k)}$. Finally, we calculate each covariance matrix via eigenvalue decomposition (Eq. \ref{eq:eigo}). 

\subsection{Second step}
The second step is similar to the first step. The main difference is the way we calculate probabilities $p_{ik}$. Instead of hard classification, in this step we utilize soft (proportional) classification. Let us focus on only one LoR with a total of $K$ components in the mixture. We want to determine $p_{ik}$ - the probability that $i$-th line belongs to $k$-th component. For each component, we calculate the line integral:
\begin{equation}
\begin{gathered}
\tilde{p}_{ik} = \int_{\gamma_i} \tau_k f_G(\bm x; \bm \mu_k, \bm \Sigma_k) d\gamma_i,
\label{pkko}
\end{gathered}
\end{equation}
where $f_G$ denoted a bivariate normal distribution and $\tau_k = \frac{L_k}{N}$ is an estimated weight of the $k$-th component. The mean vector $\bm\mu_k$ and covariance matrix $\bm\Sigma_k$ are known from the previous step. Mapping $\gamma_i$ is a parametrization of the line of response we are currently observing. 
The solution to this integral is a point from the univariate normal distribution (see \ref{AppA} and Eq. \ref{eq:pdf_1d}). 
Keep in mind that, since $i$-th response have occurred, the overall probability must be 1. Therefore, we define
\begin{equation}
\begin{gathered}
p_{ik} = \frac{\tilde{p}_{ik}}{\sum_{k=1}^K \tilde{p}_{ik}}.
\label{pks}
\end{gathered}
\end{equation}
We keep the same ratio as in $\tilde{p}_{ik}$, but force the probabilities to add up to one for each LoR. Additionally, the estimation of the number of LoR-s from Eq. \ref{eq:estim_num_of_lines} still holds.

After we updated all membership probabilities $p_{ik}$, we calculate a new mean vector $\bm\mu_k$ as already described (Eq. \ref{eq:solve_mu}). With newly estimated mean vectors, we determine new covariance matrices $\bm\Sigma_k$ in the same way as presented in the last subsection. Then we repeat the procedure, i.e. we calculate $p_{ik}$ from $\bm\mu_k$ and $\bm\Sigma_k$. This can be described in 8 steps:
\begin{enumerate}
\item Calculate $p_{ik}$ (Eq. \ref{pkko}  and Eq. \ref{pks}) for all lines $i$ and all components $k$ that best fit the current model.
\item Estimate the mean vector for each component according to newly obtained $p_{ik}$.
\item Calculate $\hat{\sigma}_1^{(k)}$ and $\hat{\sigma}_2^{(k)}$ from Eq. \ref{eq:moment_sol}.
\item Determine $\hat{\varphi}_0^{(k)}$ from Eq. \ref{eq:system_phi} and \ref{eq:system_phi_poly}.
\item Reevaluate $\hat{\sigma}_1^{(k)}$ and $\hat{\sigma}_2^{(k)}$ from Eq. \ref{eq:system_sigmas}.
\item Update $\hat{\varphi}_0^{(k)}$ from Eq. \ref{eq:system_phi} and \ref{eq:system_phi_poly} with new $\hat{\sigma}_1^{(k)}$ and $\hat{\sigma}_2^{(k)}$.
\item Use eigenvalue decomposition to calculate the covariance matrix (Eq. \ref{eq:eigo}).
\item Do 3)-7) for every component in the mixture ($k=1,2,3,..,K$).
\item Repeat 1)-8) until the change in estimated weight of all components is sufficiently small.
\end{enumerate}

We obtained all mixture parameters: $\bm\mu_k$ and $\bm\Sigma_k$ and $\tau_k$. Such a set of parameters represents continuous model of the reconstructed image. Pay attention that the proposed algorithm is numerically efficient, in spite of the number of equations in the previous section.\footnote{Implementation of described algorithm is available on \url{https://github.com/tm2005/Accurate-PET-Reconstruction-from-Reduced-Set-of-Measurements-based-on-GMM}.}

\section{Results}
Our test mixture consists of three Gaussian components, in which two components highly overlap. 

\begin{figure}
\centering
 \includegraphics[width=0.55\linewidth]{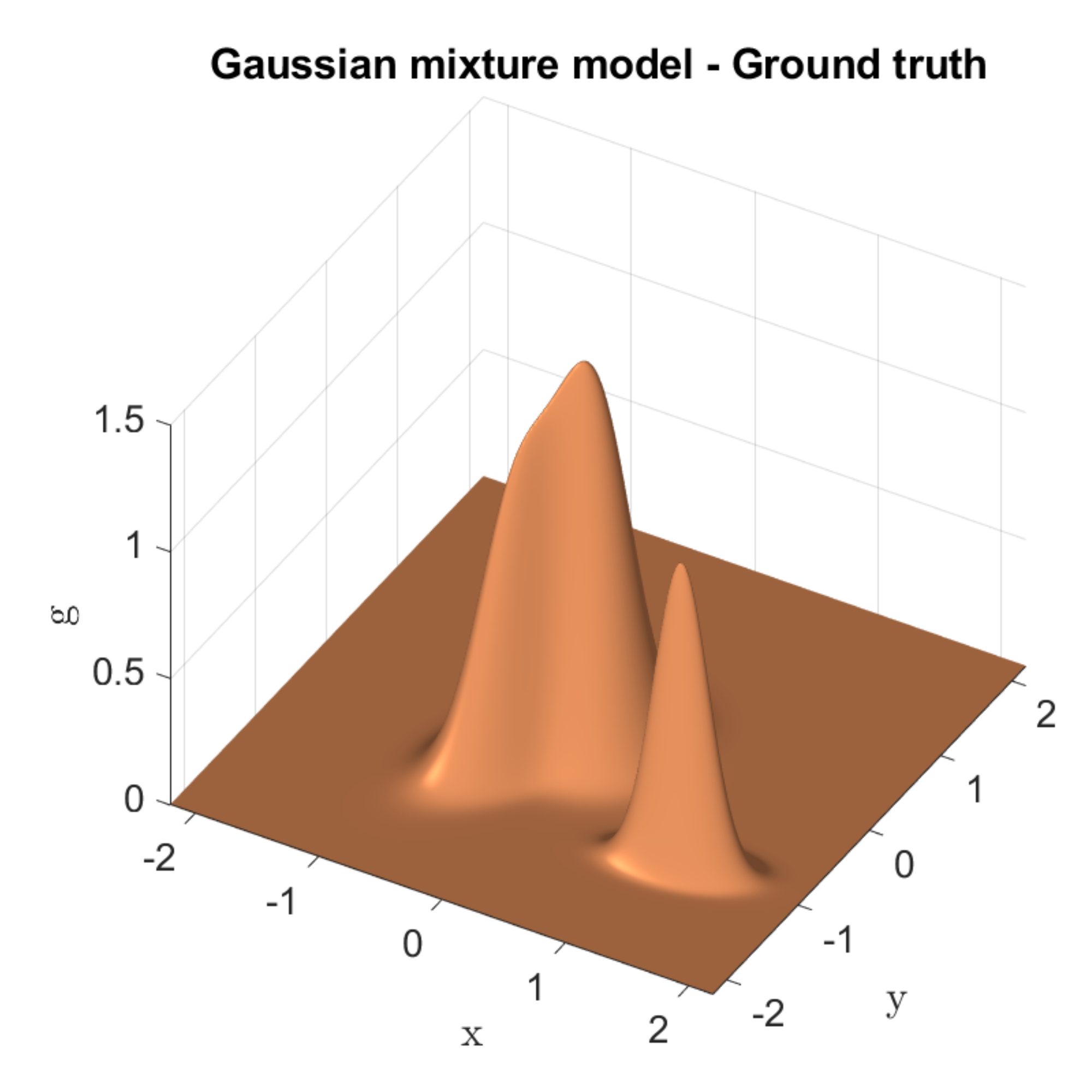}
\caption{The probability density function of the GMM (Ground truth)} 
\label{fig:gmm_GT}
\end{figure}

The first component has mean vector $\bm\mu_1=\begin{pmatrix}0\\0 \end{pmatrix}$ and covariance matrix $\bm\Sigma_1=\begin{pmatrix}0.0625&0\\0&0.0625 \end{pmatrix}$. The second component highly overlaps with the first component and its parameters are $\bm\mu_2=\begin{pmatrix}-0.4\\-0.4 \end{pmatrix}$ and $\bm\Sigma_2=\begin{pmatrix}0.04&0.03\\0.03&0.09 \end{pmatrix}$. The third component has the mean $\bm\mu_3=\begin{pmatrix}1.25\\-1 \end{pmatrix}$ and covariance $\bm\Sigma_3=\begin{pmatrix}0.04&0.006\\0.006&0.01 \end{pmatrix}$. During the simulation, we created $N_1 = 3500$ lines that originate from the first component, $N_2 = 2500$ lines that originate from the second component, and $N_3 = 1000$ from the last component. Hence, the corresponding weights in the Gaussian mixture model are $\tau_1 = 0.5$, $\tau_2 = \frac{5}{14} \approx 0.36$, and $\tau_3 = \frac{1}{7} \approx 0.14$. The PDF of our GMM is (as depicted in Fig \ref{fig:gmm_GT}):

\begin{equation}
\begin{gathered}
g(x,y) = \tau_1 \cdot f_G(x,y;\bm\mu_1,\bm\Sigma_1) + \tau_2 \cdot f_G(x,y;\bm\mu_2,\bm\Sigma_2) \\+\tau_3 \cdot f_G(x,y;\bm\mu_3,\bm\Sigma_3),
\end{gathered}
\label{eq:example_ideal}
\end{equation} where $f_G$ is bivariate normal distribution as in Eq \ref{eq:multivar_gauss}.

We have repeated the simulation $n=100$ times with the same parameters and applied the proposed algorithm to determine the unknown parameters of the Gaussian mixture model. To check our algorithm, we measure four errors. The first one measures an error between the ideal and estimated mean vector
\begin{equation}
\bm\mu_{er_i} = ||\bm\hat{\bm\mu}_i -\bm\mu_i||_2.
\end{equation}
The second one evaluates the error between the ideal and estimated covariance matrix
\begin{equation}
\bm\Sigma_{er_i} = ||\bm\hat{\bm\Sigma}_i - \bm\Sigma_i||_F.
\end{equation}
Also, we measure the error between the ideal and estimated weights in the GMM 
\begin{equation}
\tau_{er_i} = |\hat{\tau}_i-\tau_i|.
\end{equation}

\begin{figure}[h]
\centering
    \begin{subfigure}{0.32\textwidth}
        \includegraphics[width=\linewidth]{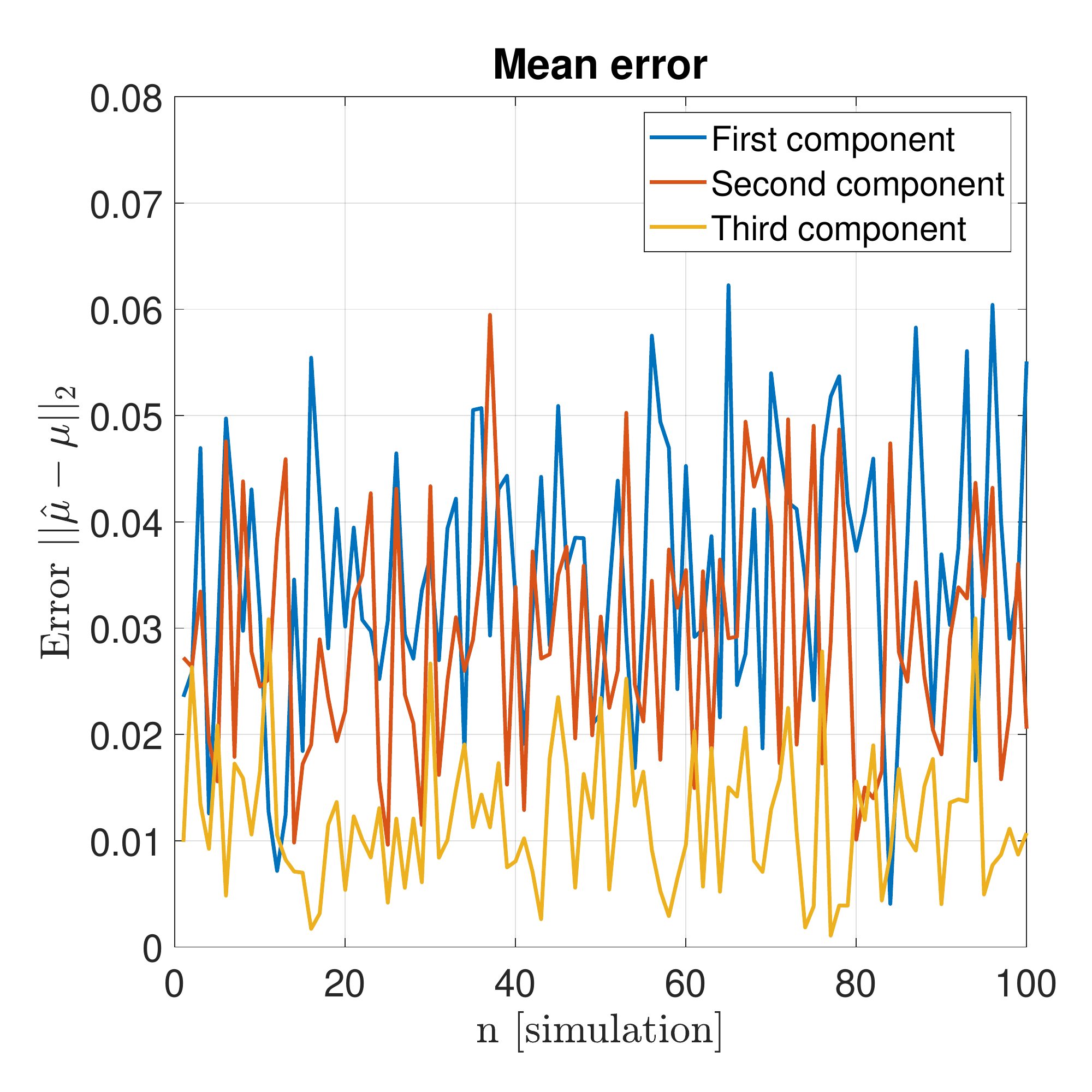}
        \caption{Mean vector error}
        \label{fig:means_error}
    \end{subfigure}\hspace*{1mm}
    \begin{subfigure}{0.32\textwidth}
        \includegraphics[width=\linewidth]{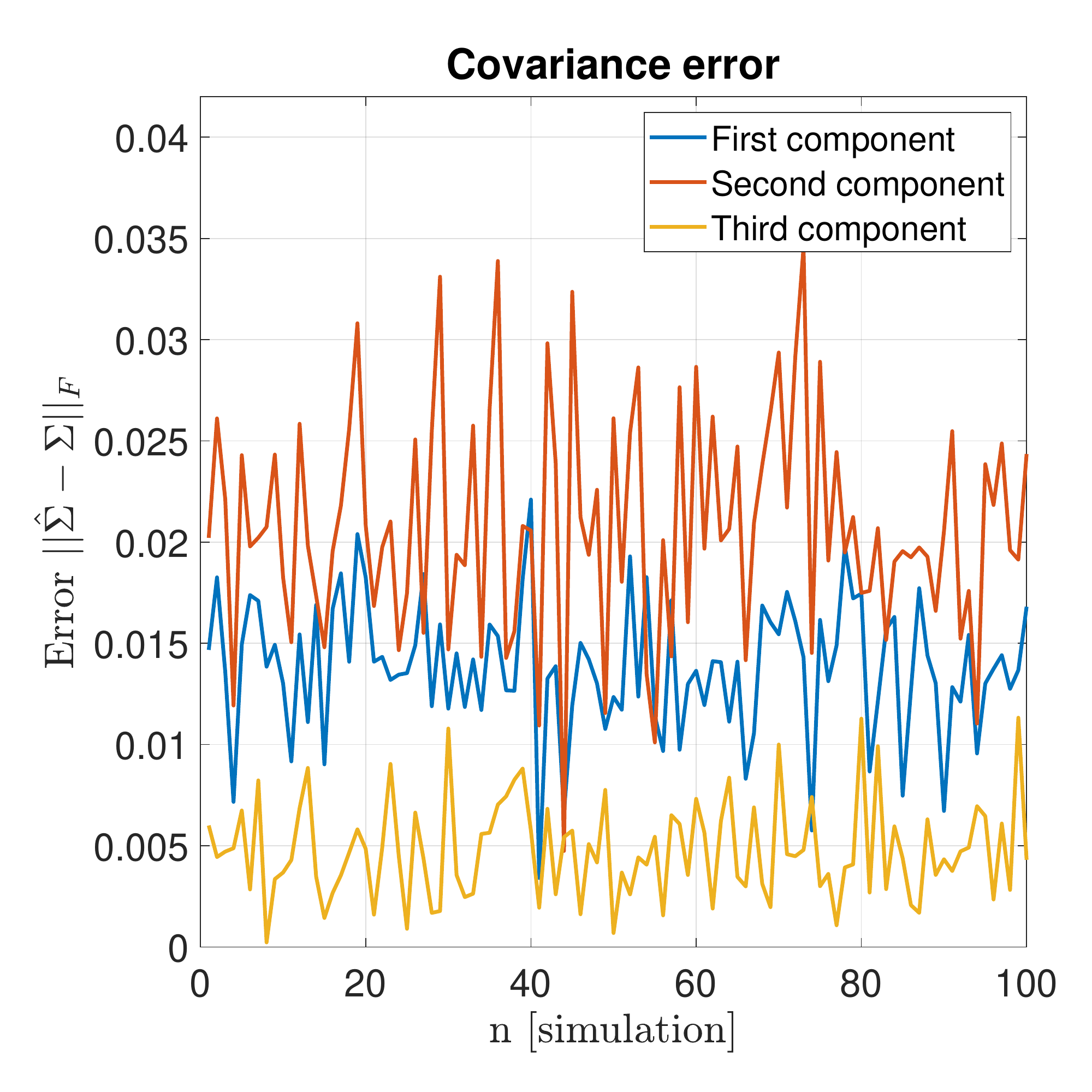}
        \caption{Covariance matrix error}
        \label{fig:cov_error}
    \end{subfigure}\hspace*{1mm}
    \begin{subfigure}{0.32\textwidth}
        \includegraphics[width=\linewidth]{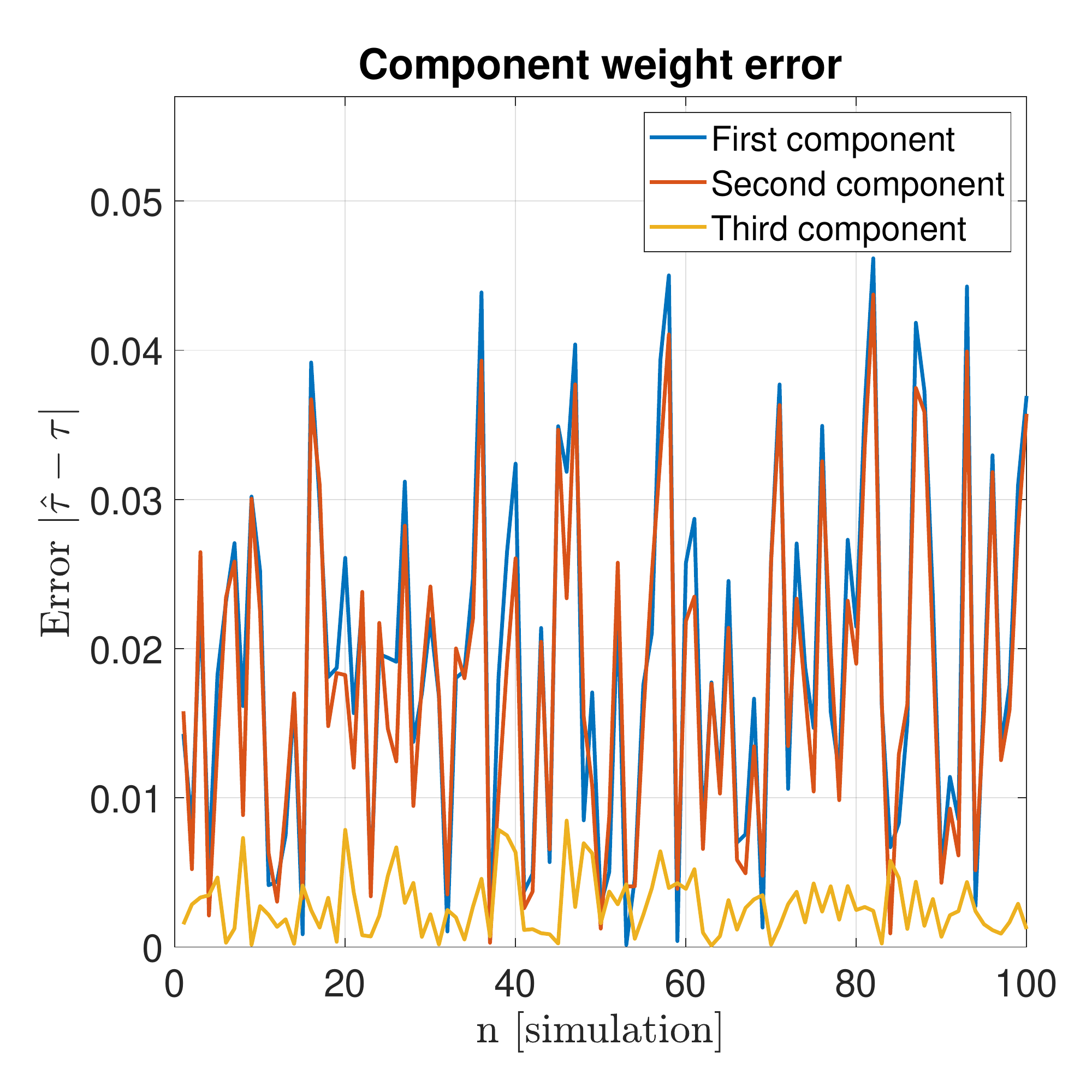}
        \caption{Mixture weight error}
        \label{fig:tau_error}
    \end{subfigure}
\caption{Errors of individual parameters of GMM in each simulation} 
\label{fig:PET_sim_many}
\end{figure}

Unlike the mentioned measures that deal with each parameter separately, the last one is the Kullback\text{-}Leibler (KL) divergence. It measures an overall error, since it compares the two PDF-s:
\begin{equation}
D_{KL}(\hat{g}|g) = \int\displaylimits_{-\infty}^{+\infty}\int\displaylimits_{-\infty}^{+\infty} \hat{g}(x,y)\log\left(\frac{\hat{g}(x,y)}{g(x,y)}\right) dxdy,
\end{equation}
where $g(x,y)$ is the ideal PDF (Eq. \ref{eq:example_ideal}) and $\hat{g}(x,y)$ is the estimate of the PDF obtained via the proposed algorithm.

In Figure \ref{fig:PET_sim_many}, we can see mean, covariance, and components weight error in each simulation. Pay attention that the non-overlapping component has smaller errors. Moreover, all errors are very small, and the averages are presented in Table \ref{tab:errors}.
\begin{table}[t]
\centering
 \begin{tabular}{|c | c | c | c|} 
 \hline
 Component  & Avg. Mean & Avg. Cov. & Avg. Weight \\ [0.5ex] 
  \hline
 First & 0.035 & 0.014 & 0.019 \\
 Second & 0.029 & 0.021 & 0.018 \\
 Third & 0.011 & 0.004 & 0.002 \\
 \hline
 \end{tabular}
 \caption{Average error of parameters of GMM in 100 simulations}
 \label{tab:errors}
\end{table}

\begin{figure}
\centering
 \includegraphics[width=0.37\linewidth]{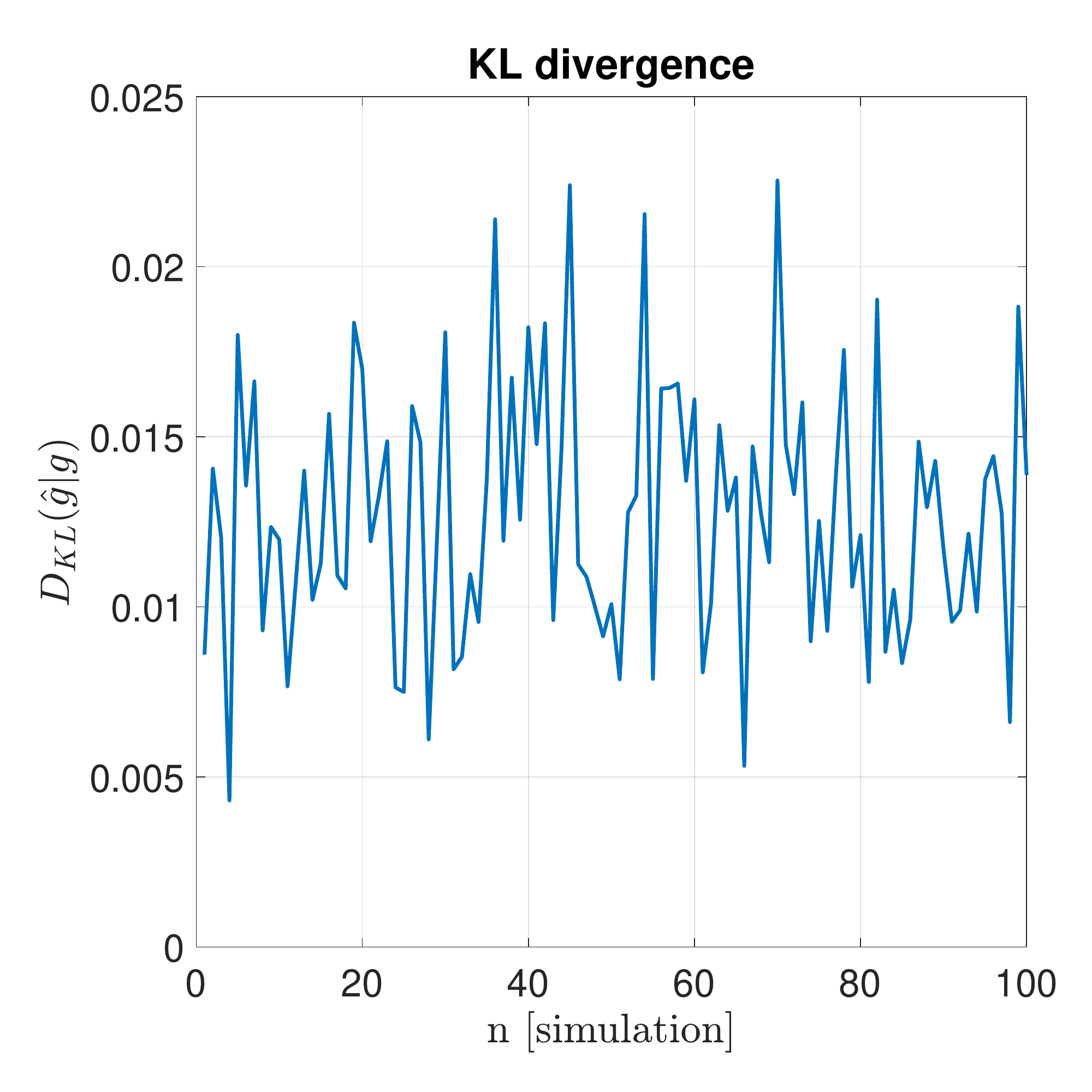}
\caption{KL divergence in each simulation} 
\label{fig:KL_error}
\end{figure}

The KL divergence of 100 simulations is depicted in Fig. \ref{fig:KL_error}. The calculation of the KL divergence is done numerically. The KL divergence is always smaller than $0.023$, while the mean value of all simulations is $0.013$.

\begin{figure}
\centering
 \includegraphics[width=0.55\linewidth]{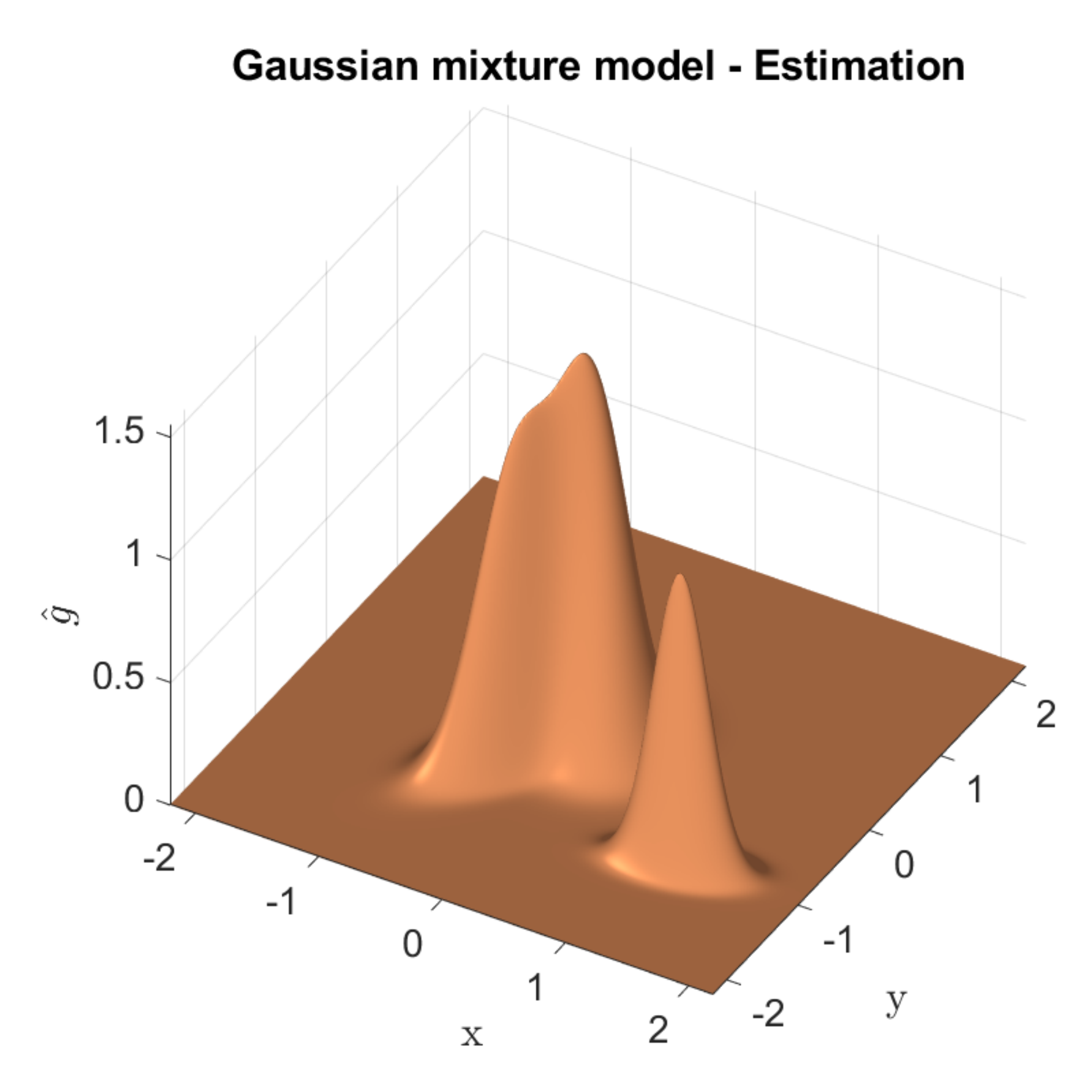}
\caption{Estimated PDF of the GMM} 
\label{fig:gmm_Estim}
\end{figure}

From all that has been said, we conclude that the estimated PDF of GMM highly resembles the true PDF.  A typical reconstruction of the observed setup can be seen in Fig. \ref{fig:gmm_Estim}. As expected, the difference between the true PDF (Fig. \ref{fig:gmm_GT}) and the estimated one is not noticeable.

\section{Conclusion}
We presented a novel method for estimation of unknown parameters of the Gaussian mixture model in Positron Emission Tomography. In contrast to competitive reconstruction methods based on pixel or voxel grid, the obtained GMM is continuous, and virtually of infinite resolution. It can be directly used for further analytical processing. There are several well-known GMM estimation methods based on samples. In PET imaging, the samples are not known. The challenge was to estimate the GMM parameters from projections, namely from the lines of response that fires from unknown radioactive decay points under some random angles. We solved the problem analytically, and proposed a systematic approach for obtaining the unknown parameters, under the assumption that the number of Gaussian components is known.

First, we estimate the mean vectors for each component in the mixture. For each component, we get a $2\times 2$ linear system of equations. Then, the covariance matrices of each component are estimated in four steps. In the first step, we propose to use higher moments of each Gaussian component to estimate variances in the principal axes. Then, we obtain the direction of the larger principal axis. In the third step, we additionally tune the variance estimates. Finally, we use the eigenvalue decomposition to get the desired covariance matrix of each component. Weights of Gaussian components are given by the membership probabilities. Described steps are integrated into an iterative ML-like algorithm.
 
The results presented in this paper show that recovery of the unknown parameters is possible even when two components significantly overlap. A relatively small number of LoR-s is needed for an accurate reconstruction, thus eventually leading to lower radiation doses in the PET imaging.  

\section*{Funding}
This research was supported by the Croatian Science Foundation [IP-2019-04-6703].

\appendix
\section{Line integration of centered and not rotated bivariate normal distribution}\label{AppA}
In this appendix we present a detailed procedure on how to calculate an integral stated in Eq. \ref{eq:normal_integrate}: 
\[
f_{p_{0_\varphi}}(s)=\int_{\gamma_{\varphi}}f_{Gc}(x,y)d\gamma_{\varphi}.
\]
We remind the reader that $|\varphi|\leq\frac{\pi}{2}$. Since we are dealing with a line integral, we calculate $||\gamma_{\varphi}'(t)||_2$:
\[
||\gamma_{\varphi}'(t)||_2 = ||(1,k)||_2 = \sqrt{1+k^2} = \frac{1}{\cos(\varphi)}.
\]

We calculate $f_{p_{0_\varphi}}(s)=\int_{\gamma_{\varphi}}f_{Gc}(x,y)d\gamma_{\varphi}$:

\[
\begin{gathered}
\int\displaylimits_{\gamma_{\varphi}} f_{Gc}(x,y)d\gamma = \int\displaylimits_{-\infty}^{+\infty}\frac{1}{2\pi\sigma_1\sigma_2}\exp(-(\frac{t^2}{\sigma_1^2}+ \frac{(kt+l)^2}{\sigma_2^2}))\cdot ||\gamma'(t)||dt=\\
\frac{\exp(-\frac{t^2}{2\sigma_2^2})}{2\pi\sigma_1\sigma_2\cos(\varphi)}\int\displaylimits_{-\infty}^{+\infty}\exp(-(t^2(\frac{1}{2\sigma_1^2} + \frac{k^2}{2\sigma_2^2}) + \frac{kl}{\sigma_2^2}t))dt = \\
\frac{\exp(-\frac{t^2}{2\sigma_2^2}) \exp(\frac{k^2t^2}{4\sigma_2^4}\frac{1}{\frac{1}{2\sigma_1^2} + \frac{k^2}{2\sigma_2^2}})}{2\pi\sigma_1\sigma_2\cos(\varphi)}\int\displaylimits_{-\infty}^{+\infty}\exp(-(t\sqrt{\frac{1}{2\sigma_1^2} + \frac{k^2}{2\sigma_2^2})} + \frac{1}{\sqrt{\frac{1}{2\sigma_1^2} + \frac{k^2}{2\sigma_2^2})}}\frac{kl}{2\sigma_2^2})^2)dt=\\
\frac{1}{2\pi\sigma_1\sigma_2\cos(\varphi)}\frac{\sqrt{\pi}}{\sqrt{\frac{1}{2\sigma_1^2} + \frac{k^2}{2\sigma_2^2}}}\exp(\frac{-l^2}{2\sigma_2^2}(1-\frac{2\sigma_1^2\sigma_2^2}{\sigma_2^2+k^2\sigma_1^2}\frac{1}{2\sigma2^2}))=\\
\frac{1}{\sqrt{2\pi}\sigma_1\sigma_2\cos(\varphi)}\frac{\sigma_1\sigma_2}{\sqrt{\sigma_2^2+k^2\sigma_1^2}}e^{\frac{-l^2}{2(\sigma_2^2+k^2\sigma_1^2)}}=\\
\frac{1}{\sqrt{2\pi}}\frac{1}{\cos(\varphi)\sqrt{\sigma_2^2+\tan^2(\varphi)\sigma_1^2}}\exp(\frac{-s^2}{2\cos^2(\varphi)(\sigma_2^2+\tan^2(\varphi)\sigma_1^2)})=\\
\frac{1}{\sqrt{2\pi}\sqrt{\sigma_2^2\cos^2(\varphi)+\sigma_1^2\sin^2(\varphi)}}\exp(\frac{-s^2}{2(\sigma_2^2\cos^2(\varphi)+\sigma_1^2\sin^2(\varphi))}) = \\
\frac{1}{\sqrt{2\pi}\sigma_{p_0}(\varphi)}\exp(-\frac{s^2}{2\sigma_{p_0}(\varphi)^2}).
\end{gathered}
\]
Here, we used known result:
\[
\int\displaylimits_{-\infty}^{+\infty}\exp(-(at+b)^2)dt = \frac{\sqrt{\pi}}{|a|},\,\,\, a\neq 0.
\]
\bibliographystyle{plain}
\bibliography{biblio.bib}
\end{document}